\documentclass[12pt]{article}
\usepackage{amsmath}
\usepackage{amssymb}
\usepackage{epsfig}
\usepackage[bf, small, hang]{caption}

\def\mycaption#1{\caption{#1}}

\usepackage{bbm}
\def\Id{\mathbbm{1}}
\def\NN{\mathbbm{N}}

\def\GA{\textsf{GA}}
\def\ssGA{\textsf{ssGA}}
\def\genGA{\textsf{genGA}}
\def\nih{\textsf{NiH}}
\def\xx{\mathbf{x}}
\def\yy{\mathbf{y}}
\def\vv{\mathbf{v}}
\def\pha{\varphi}
\def\Perm{\mathfrak{S}}

\begin{document}

\title{\large\sc Genetic Algorithms in Time-Dependent Environments}

\author{\normalsize Christopher Ronnewinkel,\\[-4pt]
\small\parbox{0.35\textwidth}{\begin{center}
{\sl temporary address:}\\[4pt]
Institut f\"ur Neuroinformatik,\\
Ruhr-Universit\"at Bochum,\\
D-44780 Bochum, Germany
\end{center}}
\and\normalsize  Claus O. Wilke and Thomas Martinetz\\
\small\parbox{0.45\textwidth}{\begin{center}
Institut f\"ur Neuro- und Bioinformatik,\\
Universit\"at L\"ubeck,\\
Ratzeburger Allee 160,\\
D-23538 L\"ubeck, Germany
\end{center}}
}
%
%

\date{\vskip-1.2em\normalsize
Contact: {\tt ronne@neuroinformatik.ruhr-uni-bochum.de}\\[1.5ex]
(to be published in the {\sl Proceedings of the 2nd EvoNet Summerschool},\\
Natural Computing Series, Springer)\\[4ex]
November 2, 1999}

\maketitle

\begin{abstract}
The influence of time-dependent fitnesses on the infinite population
dynamics of simple genetic algorithms (without crossover) is analyzed. 
Based on general arguments, a schematic phase diagram is constructed
that allows one to characterize the asymptotic states in dependence on the
mutation rate and the time scale of changes.
Furthermore, the notion of {\it regular\/} changes is raised for which
the population can be shown to converge towards a {\it generalized\/}
quasispecies. Based on this, error thresholds and an optimal mutation
rate are approximately calculated for a generational genetic algorithm
with a moving needle-in-the-haystack landscape. The so found phase
diagram is fully consistent with our general considerations.
\end{abstract}

Genetic algorithms (\GA{}s) as special instances of evolutionary
algorithms have been established during the last three decades as
optimization procedures, but mostly for
static problems (see \cite{baeck} for an overview and \cite{baeckhb}
for an in-depth presentation of the field).
In view of real-world applications, such as routing
in data-nets, scheduling, robotics etc., which include
essentially dynamic optimization problems, there are two alternative
optimization strategies. On the one hand, one can take snapshots of
the system and search ``offline'' for the optimal solutions of the static
situation represented by each of these snapshots. In this approach,
the algorithm is restarted for every snapshot and solves the new
problem from scratch. On the other hand,
the optimization algorithm might reevaluate the real, current
situation in order to reuse information gained in the past.
In this case, the algorithm works ``online''. As can be argued from the
analogies to natural evolution, evolutionary
algorithms seem to be promising candidates for ``online'' optimization
\cite{baeck, brankereview}.
The reevaluation of the situation or environment then introduces a 
{\it time-dependency\/} of the fitness landscape. This time-dependency occurs as
external to the algorithm's population and does not emerge from coevolutive
interactions. Coevolutive interactions as an alternative source of
time dependency in the fitness landscape are not within the scope of this work.  

In the last years, many different methods and extensions of
standard evolutionary algorithms for the case of time-dependent
fitnesses have been analyzed on the basis of experiments (see
\cite{brankereview} for a review) but only seldom on the basis of
{\it theoretical\/} arguments (see \cite{rowegecco, nehaniv}).
To take a step into the direction of a better theoretical
understanding of ``online'' evolutionary algorithms,
we will study the effects of simple time dependencies
of the fitness landscape on the dynamics of \GA{}s (without crossover), or more
generally saying, of populations under mutation and probabilistic
selection. As we will see, it is possible to characterize the
asymptotic states of such a system for a particular class of dynamic
fitness landscapes that is introduced below.
The asymptotic state forms the basis on which it can be decided
whether the population is able to adapt to, or track, the changes in
the fitness landscape. Our mathematical formalism applies to \GA{}s as well as
to biological self-replicating systems, since the analyzed \GA{} model
and Eigen's quasispecies model \cite{eigen71, eigen79, eigen89} in the
molecular evolution theory (see \cite{baake} for a recent review) are very
similar. Hence, all introduced
concepts for \GA{}s are valid and relevant in
analogous form for molecular evolutionary systems.

In the following section, we will introduce the model to be analyzed
and show the correspondence to the quasispecies model. Then, we will
introduce the mathematical framework, based on which we will formally characterize the
asymptotic state as fixed point. After presenting the main concepts,
we will proceed with the construction of a phase diagram that allows
to characterize the order found in the asymptotic state for different
parameter settings. Finally, a moving
needle-in-the-haystack (\nih{}) landscape is analyzed and its phase diagram,
including the optimal mutation rate, is calculated.

\section{Mathematical Framework}
In order to study the influence of a time-dependent
fitness landscape on the dynamics of a genetic algorithm (\GA{}), we
consider \GA{}s to be discrete dynamical systems. A detailed
introduction to the resulting dynamical systems model is given by
Rowe \cite{rowe} (in this book). Here, we will only shortly
introduce the basic concepts and the notations we use within the
present work.

The \GA{} is represented as a generation operator 
$\smash{G_{t}^{(m)}}$ acting on the space
$\Lambda_{m}$ of all populations of size $m$ for some given encoding
of the population members. If we choose
the members $i$ to be encoded as bit-strings of
length $l$, this state space is given by
$$\Lambda_{m}=\{(n_{0},\ldots,n_{2^{l}-1})/m \mid \textstyle\sum_{i}n_{i}=m,
n_{i}\in\NN_{0}\},$$
where $n_{i}$ denotes the number of bit-strings in the population, 
which are equal to the binary representation of $i\in\{0,\ldots,2^{l}-1\}$.

The generation operator maps the present population onto the next generation,
\begin{equation*}
\xx(t+1)=G_{t}^{(m)}[\xx(t)].
\end{equation*}
This is achieved by applying a sampling procedure that draws the
members of the next generation's population $\xx(t+1)$ according to
their expected concentrations $\langle
\xx(t+1)\rangle\in\Lambda_{\infty}$ which are defined by the mixing \cite{rowe,vose}
and the selection scheme. For an infinite population size, the sampling
acts like the identity resulting in
\begin{equation*}
G_{t}^{(\infty)}\xx(t)=\xx(t+1)=\langle\xx(t+1)\rangle.
\end{equation*}
Hence, $G_{t}:=G_{t}^{(\infty)}$ represents in fact the mixing and
selection scheme. For finite population size,
$\langle\xx(t+1)\rangle\in\Lambda_{\infty}$ is approximated by using the
sampling process to obtain $\xx(t+1)\in\Lambda_{m}$. The deviations 
thereby possible become larger with decreasing $m$ and distort the finite
population dynamics as compared to the infinite population case. This
results in fluctuations and epoch formation as shown in \cite{rowe,vose,nimw97}.
In the following, we will consider the infinite population limit,
because it reflects the exact flow of probabilities for a particular
fitness landscape. In a second step, the fluctuations and epoch formation
introduced by the finiteness of a real population can be studied on
the basis of that underlying probability flow.

The generation operator is assumed to decompose into a separate
mutation and a separate selection operator, like
\begin{equation}
G_{t}=M\cdot S(t),
\label{eq:genga}\end{equation}
where the selection operator $S(t)$ contains the time dependency of the
fitness landscape. Crossover is not considered in this work.

Inspired by molecular evolution, and also by common usage, we
assume that the mutation acts like flipping each bit with probability $\mu$.
If we set the duration of one generation to $1$, $\mu$ equals to the mutation rate. The
mutation operator then takes on the form
$$M=\left(\begin{matrix}1-\mu &\mu\\\mu & 1-\mu\end{matrix}
\right)^{\otimes l}\mbox{,\quad i.\ e.}\quad
M_{ij}=\mu^{d_{\rm H}(i,j)}(1-\mu)^{l-d_{\rm H}(i,j)},$$
where $\otimes$ denotes the Kronecker (or canonical tensor) product
and $d_{\rm H}(i,j)$ denotes the Hamming distance of $i$ and $j$.

To keep the description analytically tractable, we will focus on
fitness-proportionate selection,
\begin{align*}
S(t)\cdot\xx=F(t)\cdot\xx\big/\langle f(t)\rangle_{\xx},\quad
\mbox{where } F(t)&={\rm diag}\big(f_{0}(t),\ldots,f_{2^{l}-1}(t)\big)\\
\mbox{ and 
$\langle f(t)\rangle_{\xx}$}&=\textstyle\sum_{i}f_{i}(t)x_{i}
=\|F(t)\cdot\xx\|_{1}.
\end{align*}
This will already provide us with some insight into the general behavior of a
\GA{} in time-dependent fitness landscapes.

Since the \GA{} corresponding to Eq.\ \ref{eq:genga} applies
mutation to the current population and selects the new population with
complete replacement of the current one, it is called a {\it generational\/} 
\GA{} (\genGA{}). In addition to \genGA{}s, {\it steady-state\/} \GA{}s (\ssGA{}s)
with a two step reproduction process are also
in common use: First, a small fraction $\gamma$ of the current population is
chosen to produce $m\gamma$ mutants according to some
heuristics. Second, another fraction $\gamma$ of the current population
is chosen to get replaced by those mutants according to some other
heuristics (see \cite{dejong,rogers,branke} and references therein).
We can include \ssGA{}s into our description in an approximate fashion by simply
bypassing a fraction $(1-\gamma)$ of the population into the selection
process without mutation, whereas the remaining fraction $\gamma$ gets
mutated before it enters the selection process. The generation operator then reads
\begin{equation}
G_{t}=\left[(1-\gamma)\Id+\gamma M\right]S(t).
\label{eq:ssga}\end{equation}
By varying $\gamma$ within the interval $]0,1]$, we can
interpolate between steady-state behavior (\ssGA{}) for $\gamma\ll 1$
and generational behavior (\genGA{}) for $\gamma=1$.
Equation \ref{eq:ssga} is only an approximation of the true generation
operator for \ssGA{}s because the heuristics involved in the choice of
the mutated and replaced members are neglected. But in the next
section, the heuristics are expected to play a minor role for our
general conclusion on an inertia of \ssGA{}s against time-variations. 

At this point, we want to review shortly the correspondence of our
\GA{} model with the quasispecies model, extensively studied by
Eigen and coworkers \cite{eigen71, eigen79, eigen89} in the context of
molecular evolution theory (see also \cite{roweqs} in this book). The quasispecies model
describes a system of self-replicating entities $i$ (e.\ g.\ RNA-, DNA-strands) with
replication rates $f_{i}$ and an imperfect copying procedure such 
that mutations occur. For simplicity reasons, the overall
concentration of molecules in the system is held constant by an excess
flow $\Phi(t)$. In the above notation, the continuous model reads
\begin{equation}
\dot{\xx}(t)=\left[M\cdot F(t)-\Phi(t)\right]\xx(t),
\label{eq:conteig}\end{equation}
where the flux needs to equal the average replication,  
$\Phi(t)=\langle f(t)\rangle_{\xx(t)}$, in order to keep the
concentration vector $\xx(t)$ normalized. This model
might then be discretized via $t\to t/\delta t$, which unveils the
similarity to a \ssGA{}:
\begin{equation}
\xx(t+1)=\left[(1-\delta t\,\langle f(t)\rangle_{\xx(t)})\Id+\delta t\,
M\cdot F(t)\right]\xx(t)\quad\mbox{for $\delta t\ll 1$.}
\label{eq:disceig}\end{equation}
By comparison with Eq.\ \ref{eq:ssga}, we can easily read off that
$\gamma=\delta t\,\langle f(t)\rangle_{\xx(t)}=:\gamma_{\xx(t)}$. This
means a low (resp.\ high) average fitness leads to a small
(resp.\ large) replacement --
a property that is not wanted in the context of optimization problems,
which \GA{}s are usually used for,
because one does not want to remain in a region of low fitness
for a long time. Another difference to \ssGA{}s is the fact that in
the continuous Eigen model, selection acts only on the mutated fraction of the
population -- although this leads only to subtle differences in the dynamics
of \ssGA{}s and the Eigen model. 

Equation \ref{eq:conteig} is commonly referred to as `continuous Eigen
model' in the literature, because of the continuous
time, and Eq.\ \ref{eq:disceig} is simply its discretized form which
can be used for numerical calculations. 
Nonetheless, the notion `discrete Eigen model' is seldom used for
Eq.\ \ref{eq:disceig} but it is often used for the \genGA{},
\begin{equation}
\xx(t+1)=\left[M\cdot S(t)\right]\xx(t),
\label{eq:geneig}\end{equation}
in the literature.
This stems from the identical asymptotic behavior of Eqs.\
\ref{eq:disceig} and \ref{eq:geneig} for static fitness
landscapes. However, there are differences for time-dependent fitness
landscapes, as we will see in the following two sections.

\section{Regular Changes and Generalized Quasispecies}\label{sec:fp}
In the case of a static landscape, the fixed points of the generation
operator, which are in fact stationary states of the evolving system
(if contained within $\Lambda_{m}$, see \cite{rowe}),
can be found by solving an eigenvalue problem, because of
\begin{equation}
\xx=G\xx\quad\Longleftrightarrow\quad MF\,\xx=\langle
f\rangle_{\xx}\xx\ .
\label{eq:statfp}\end{equation}
Let $\lambda_{i}$ and $\vv_{i}$ denote the eigenvalues and eigenvectors of $MF$
with descending order $\lambda_{0}\ge\cdots\ge\lambda_{2^{l}-1}$ and $\|\vv_{i}\|_{1}=1$.
For $\mu\not=0,1$ the Perron-Frobenius theorem assures the
non-degeneracy of the eigenvector $\vv_{0}$ to the largest eigenvalue and moreover
it assures $\vv_{0}\in\Lambda_{\infty}$. Often, $\vv_{0}$ is called
Perron vector. After a transformation to the basis of the
eigenvectors $\{\vv_{i}\}$ it can be straightforwardly shown that $\xx(t)$
converges to $\vv_{0}$ for $t\to\infty$. The population represented by
$\vv_{0}$ was called the `quasispecies' by Eigen, because this
population does not consist of only a single dominant genotype, or
string, but it consists of a particular stable mixture of different
genotypes.

Let us now consider time-dependent landscapes. If the time dependency
is introduced simply by a single scalar factor, like
$$F(t)=F\,\rho(t)\quad\mbox{with $\rho(t)\ge0$ for all $t$,}$$
it immediately drops out of the selection operator for \GA{}s. For the
continuous Eigen model, we note that the eigenvectors of $F(t)$ 
and $F$ are the same and that
$\lambda_{i}(t)=\lambda_{i}\,\rho(t)$. Since $\rho(t)\ge0$, which is
necessary to keep the fitness values positive, the order
of the eigenvalues remains, such that $MF(t)$ will show the same
quasispecies $\vv_{0}$ as $MF$. Contrasting to that special case, a
general, individual time dependency of the string's fitnesses does
indeed change the eigenvalues and eigenvectors of $MF(t)$ compared to
$MF$. For an arbitrary
time dependency the Perron vector is constantly changing, and
therefore, we cannot even define a unique asymptotic
state. However, this problem disappears for what we call {\it regular\/}
changes. After having established a theory for such changes, we can
then take into account more and more non-regular ingredients. What do we mean
by ``{\it regular\/} change''? 
We define it heuristically in the following way: a regular
change is a change that happens with fixed duration $\tau$ and obeys some
deterministic rule that is the same for all change cycles.
Let us express the latter more formally and make it more clear what
we mean by ``same rule of change''. Within a change cycle, we allow
for an arbitrary time dependency of the fitness, up to the restriction
that two different change cycles must be connected by a
permutation of the sequence space. Thus, if the time dependency is
chosen for one change cycle, e.\ g.\ the first change cycle starting
at $t=0$, it is already fixed for all other cycles, apart from the permutations. We
will represent permutations $\pi$ from the permutation group $\Perm_{2^{l}}$
of the sequence space as matrices
\begin{equation*}
(P_{\pi})_{ij}=\delta_{\pi(i),j}\quad\mbox{for $i,j\in\{0,\ldots,2^{l}-1\}$.}
\end{equation*}
The permutations of vectors $\xx$ and matrices $A$ are obtained by
\begin{gather*}
(P_{\pi}\xx)_{i}=x_{\pi(i)}\quad\mbox{and}\quad
(P_{\pi}AP_{\pi}^{\rm T})_{i,j}=A_{\pi(i),\pi(j)},
\end{gather*}
where $P_{\pi}^{\rm T}$ denotes the transpose of $P_{\pi}$ with the
property $P_{\pi}^{\rm T}=P_{\pi^{-1}}=P_{\pi}^{-1}$.

In reference to the first change cycle, we define the fitness
landscape $F(t)$ as being {\it single-time-dependent\/}, if and only if for each change
cycle $n\in\NN_{0}$ there exists a permutation
$\pi_{n}\in\Perm_{2^{l}}$, such that for all cycle phases
$\pha\in\{0,\dots,\tau-1\}$
\begin{equation*}
P_{n}\,F(\pha+n\tau)\,P_{n}^{\rm T}=
F(\pha)\qquad\mbox{(abbreviatory $P_{n}:=P_{\pi_{n}}$).}
\end{equation*}
We will call each permutation $P_{n}$ a {\it jump-rule}, or simply
{\it rule}, which connects $F(\pha+n\tau)$ and $F(\pha)$.
To make predictions about the asymptotic state of the system, we
need to relate the generation operators of different change cycles to
each other. This is readily achieved if the permutations $P_{n}$ 
{\it commute\/} with the mutation operator $M$. The condition for
this being the case is that for all $i,j$,
\begin{equation*}
M_{ij}=M_{\pi_{n}(i),\pi_{n}(j)}\quad\mbox{or equivalently}\quad
d_{\rm H}(i,j)=d_{\rm H}\big(\pi_{n}(i),\pi_{n}(j)\big).
\end{equation*}
Thus, the Hamming distances $d_{\rm H}(i,j)$ need to be {\it invariant\/}
under the permutations $P_{n}$. Geometrically this means that the
fitness landscape gets ``translated'' or ``rotated'' by those
permutations without changing the neighborhood relations. Then, we
find for arbitrary $n\in\NN$ and $\pha\in\{0,\ldots,\tau-1\}$,
\begin{equation}
G_{\pha+n\tau}=P_{n}^{\rm T}G_{\pha}P_{n}.
\label{eq:gperm}\end{equation}
To study the asymptotic behavior of the system, it is useful to
accumulate the time dependency of a change cycle by introducing
the $\tau$-generation operators,
\begin{equation*}
\Gamma_{n}:=G_{\tau-1+n\tau}\cdots G_{n\tau}\quad
\mbox{for all $n\in\NN_{0}$}.
\end{equation*}
Because of Eq.\ \ref{eq:gperm}, all these operators are related to $\Gamma_{0}$ by
\begin{equation*}
\Gamma_{n}=P_{n}^{\rm T}\Gamma_{0}P_{n},
\end{equation*}
This property allows us to write the time evolution of the system in the form
\begin{equation}
\xx(\pha+n\tau)=P_{n-1}^{\rm T}\Gamma_{0}P_{n-1}\,\,\cdots\,\,
P_{1}^{\rm T}\Gamma_{0}P_{1}\,\Gamma_{0}\,\,\xx(\pha),
\label{eq:gtev}\end{equation}
where $\pha\in\{0,\ldots,\tau-1\}$ denotes in the following always the
phase within a cycle.

Let us consider the special case of a single rule $P$ being applied at
the end of each change cycle, which results in $P_{n}=(P)^{n}$, e.\
g.\ imagine a fitness peak that moves at a constant ``velocity''
through the string space. 
We will see below that for those cases it is possible to
identify the asymptotic state with a quasispecies in analogy
to static fitness landscapes. Because of that, we can now define
the notion of {\it regularity\/} of a fitness landscape formally in the
following manner: 

A time-dependent fitness landscape $F(t)$ is {\it regular\/}, if and only if: 
(i) the fitness landscape is {\it single-time-dependent\/}, 
(ii) there exists some rule $P\in\Perm_{2^{l}}$ which is applied at
the end of each cycle such that $P_{n}=(P)^{n}$, and
(iii) the rule $P$ {\it commutes\/} with the mutation operator $M$.

In this case, we get with $P P^{\rm T}=\Id$ the time evolution
\begin{equation}
\xx(\pha+n\tau)=\big(P^{\rm T}\big)^{n}\big(P\Gamma_{0}\big)^{n}\,\xx(\pha).
\label{eq:xevol}\end{equation}
To proceed, it is useful to permute the concentrations compatible
to the rule of the fitness landscape. By this, concentrations are measured in
reference to the fitness landscape structure of the start cycle
$n=0$. 
We will denote those concentrations by $\xx'(t)$ and they are
related to the concentrations $\xx(t)$ by
\begin{equation}\begin{aligned}
\xx'(\pha+n\tau)&=(P)^{n}\,\xx(\pha+n\tau)\\
&=(P\Gamma_{0})^{n}\,\xx(\pha)
\qquad\mbox{and}\quad \xx'(\pha)=\xx(\pha).\end{aligned}
\label{eq:xxp}\end{equation} 
For example, if there is no time-dependency within the cycles,
some $x'_{i}$ will for all cycles measure the
concentration of the highest fitness string, independent of its current
position in string space. Thus, $\xx'(t)$ evolves in a fitness landscape
with periodic change, 
which can also be seen from the second line of Eq.\ \ref{eq:xxp}. 
In analogy to the static case Eq.\ \ref{eq:statfp}, the calculation 
of fixed points of $\xx'(t)$ is equivalent to an eigenvalue problem,
\begin{equation*}
\xx'(t+\tau)=\xx'(t)\quad\Longleftrightarrow\quad
P\widetilde{\Gamma}_{0}\,\xx'(t)=
\|P\widetilde{\Gamma}_{0}\,\xx'(t)\|_{1}\,\,\xx'(t),
\end{equation*}
where $\widetilde{\Gamma}_{0}$ is the {\it unnormalized\/}
$\tau$-generation operator obtained from the accumulation of the 
{\it unnormalized\/} generation operators
$\widetilde{G}_{\pha}=MF(\pha)$.

The corresponding periodic quasispecies $\vv_{0}$ can be calculated
for all phases $\pha$ of the change cycle from the Perron vector
$\vv_{0}$ of $P\Gamma_{0}$ in the following way,
\begin{equation}
\xx'(\pha+n\tau)\xrightarrow{n\to\infty}\vv_{0}(\pha)=
G_{\pha-1}\cdots G_{0}\,\vv_{0}\quad\mbox{for $\pha\in\{0,\ldots,\tau-1\}$}.
\label{eq:pqs}\end{equation}
To find the asymptotic states of the concentrations $\xx(t)$, we simply
need to invert Eq.\ \ref{eq:xxp},
\begin{equation}
\xx(\pha+\nu\tau)=\big(P^{\rm T}\big)^{\nu}\xx'(\pha+\nu\tau)\quad
\mbox{for $\nu\in\{0,\ldots, \eta-1\}$},
\label{eq:xqs}\end{equation}
where $\eta:={\rm ord}\, P$ is the order of the group element
$P\in\Perm_{2^{l}}$.

The essential reason for the existence of asymptotic states for
$\xx(t)$ lies in the finiteness of the permutation group
$\Perm_{2^{l}}$. Because of $P^{\eta}=\Id$, we find directly from
Eq.\ \ref{eq:xevol} the asymptotic state
\begin{equation*}
\xx(\pha+\tilde{n}\eta\,\tau)=(P\Gamma_{0})^{\eta\,\tilde{n}}\,\xx(t)
\xrightarrow{\tilde{n}\to\infty}\vv_{0}(\pha),
\end{equation*}
where $\vv_{0}(\pha)$ is the same as in Eq.\ \ref{eq:pqs}, because
$(P\Gamma_{0})^{\eta}$ and $P\Gamma_{0}$ have the same
eigenvectors, in particular the same Perron vector. Moreover, we get
\begin{equation}
\xx\big(\pha+(\nu+\tilde{n}\eta)\tau\big)\xrightarrow{\tilde{n}\to\infty}
\big(P^{\rm T}\big)^{\nu}\vv_{0}(\pha)\quad\mbox{for $\nu\in\{0,\ldots,\eta-1\}$},
\label{eq:pqsx}\end{equation}
which is the same result as Eqs.\ \ref{eq:pqs} and \ref{eq:xqs}
yield. In the limit of long strings $l\to\infty$, ${\rm ord}\,P$ is
not necessarily finite anymore. If ${\rm
  ord}\,P\smash{\xrightarrow{l\to\infty}}\infty$, then the asymptotic states
Eq.\ \ref{eq:pqsx} for $\xx(t)$ do not exist, but Eq.\ \ref{eq:pqs}
still holds. Hence, a quasispecies exists even in the limit
$l\to\infty$ if measured in reference to the structure of the fitness landscape.

In conclusion, Eqs.\ \ref{eq:pqs} and \ref{eq:pqsx}
represent the {\it generalized\/} quasispecies for the class of {\it regular\/} fitness
landscapes which includes as special cases static and periodic fitness
landscapes. In fact, the simplest case of a {\it regular\/} change 
is a periodic variation of the
fitness values $f_{i}(t)=f_{i}(t+\tau)$ because {\it no\/} permutations
are involved ($P=\Id$) and hence $\xx'(t)=\xx(t)$ for all $t$. The
quasispecies was generalized for this case already in \cite{wilke99a} and --
using a slightly different formalism -- in \cite{rowegecco}. 
In Section \ref{sec:genga}, we will study a more complicated example.

\section{Schematic Phase Diagram}\label{sec:spd}
\begin{figure}
\begin{center}
\ \hskip-0.07\textwidth\includegraphics[width=.9\textwidth]{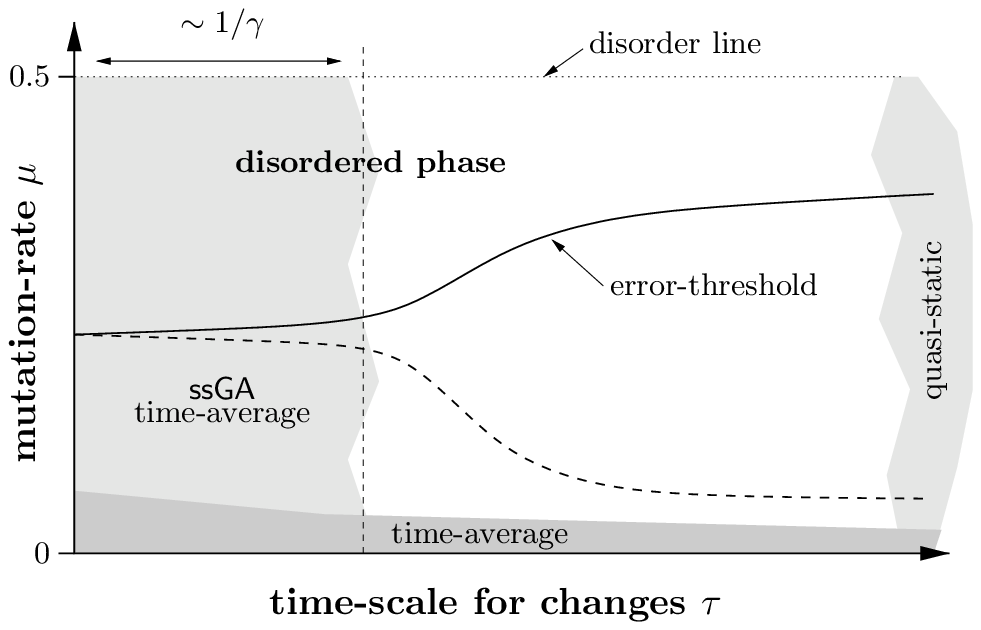}
\mycaption{%
Schematic phase diagram: time-average
regions due to low mutation (dark gray) and large inertia (light
gray, left), quasi-static region for slow changes (light gray,
right).\label{fig:phase}}
\end{center}
\end{figure}

To get an intuitive feeling for the typical behavior of \ssGA{}s and
\genGA{}s, let us consider some special lines in the plane spanned by
the mutation rate $\mu$ and the time scale for changes $\tau$, as shown in
Fig.\ \ref{fig:phase}. The mutation operator represents only for $\mu<1/2$
a copying procedure with occurring errors, whereas for $\mu>1/2$ it
systematically tends to invert strings, i.\ e.\ it resembles an
inverter with occurring errors. Since mutation should introduce 
{\it weak} modifications to the strings, we will consider only $\mu\le1/2$.

\begin{description}
\item[Disorder line:] For $\mu=1/2$, the Perron vector of $MF(t)$ is
  always $\vv_{0}^{T}=(1,\ldots,1)/2^{l}$. The population will
  therefore converge towards the disordered state. Because of the
  continuity of $M$ in $\mu$, we already enter a disordered phase
  for $\mu\approx1/2$.

\item[Time-average region:] For $\mu=0$, the mutation operator is the
  identity. We find as time evolution simply the product average over
  the fitness of the evolved time steps:
  \begin{align*}
    \xx(t+\tau)&=\left[\prod_{\pha=t}^{t+\tau-1}S(\pha)\right]\xx(t)\\
    &=\tilde{F}(t+\tau,t)\,\xx(t)\big/\|\ldots\|_{1}\mbox{\ with
      $\tilde{F}(t+\tau,t)=\prod_{\pha=t}^{t+\tau-1}F(\pha)$}.
  \end{align*}
  Since diagonal operators commute, the order in which the
  $F(\pha)$ get multiplicated does not make any difference. For the case
  of a $\tau$-periodic landscape,
  $\tilde{F}=\tilde{F}(t+\tau,t)=\tilde{F}(\tau,0)$ is {\it
  independent\/} of $t$. The quasispecies is then a linear
  superposition of the eigenvectors of the largest eigenvalue of the
  product averaged fitness landscape $\tilde{F}$ -- there might be more then one such
  eigenvector, since $\tilde{F}$ is diagonal and the
  Perron-Frobenius theorem does not apply. Because of the continuity
  of $M$ in $\mu$ the dynamics are governed already for $0<\mu\ll 1$ by the
  product average $\tilde{F}$. Analogous conclusions apply to those
  non-periodic landscapes for which by choosing a suitable time scale $\tau$
  a meaningful average $\tilde{F}(t+\tau,t)$ can be defined.
  
  For \ssGA{}s, $\gamma$ is small and we find to first order in
  $\tau\gamma$:
  \begin{multline*}\xx(t+\tau)=(1-\tau\gamma)\tilde{F}(t+\tau,t)\\
    +\tau\gamma\biggl(\frac{1}{\tau}\sum_{\pha=0}^{\tau-1}S(t+\tau)\cdots
    \underbrace{M}_{\hbox to 1pt {\scriptsize $\pha$th factor from left}}
\cdots S(t)\biggr)
    +\mathcal{O}\big((\tau\gamma)^{2}\big).
  \end{multline*}
  If $\tau\gamma\ll1$ holds, the time evolution is governed by
  $\tilde{F}(t,t+\tau)$. For changes on a time scale $\tau$, we find
  time-averaged behavior if $\tau\ll 1/\gamma$. Thus, the width of the
  time-average region is proportional to $1/\gamma$. A detailed analysis
  of the effect of the different positions of the mutation operator $M$
  within the $\tau\gamma$-term, which is
  otherwise an arithmetic time-average, has not yet been carried out.
  
\item[Quasi-static region:] If the changes happen on a time scale $\tau$
  very large compared to the average relaxation time ($\sim
  1/\langle\lambda_{0}-\lambda_{1}\rangle$) the quasispecies
  grows nearly without noticing the changes. Thus, in the
  quasi-static region all quasispecies that might be expected from
  the static landscapes $\tilde{F}=F(t)$ will occur at some
  time during one cycle $\tau$.
\end{description}

\begin{figure}
  \begin{center}
    \ \hskip-.12\textwidth\includegraphics[width=.9\textwidth]{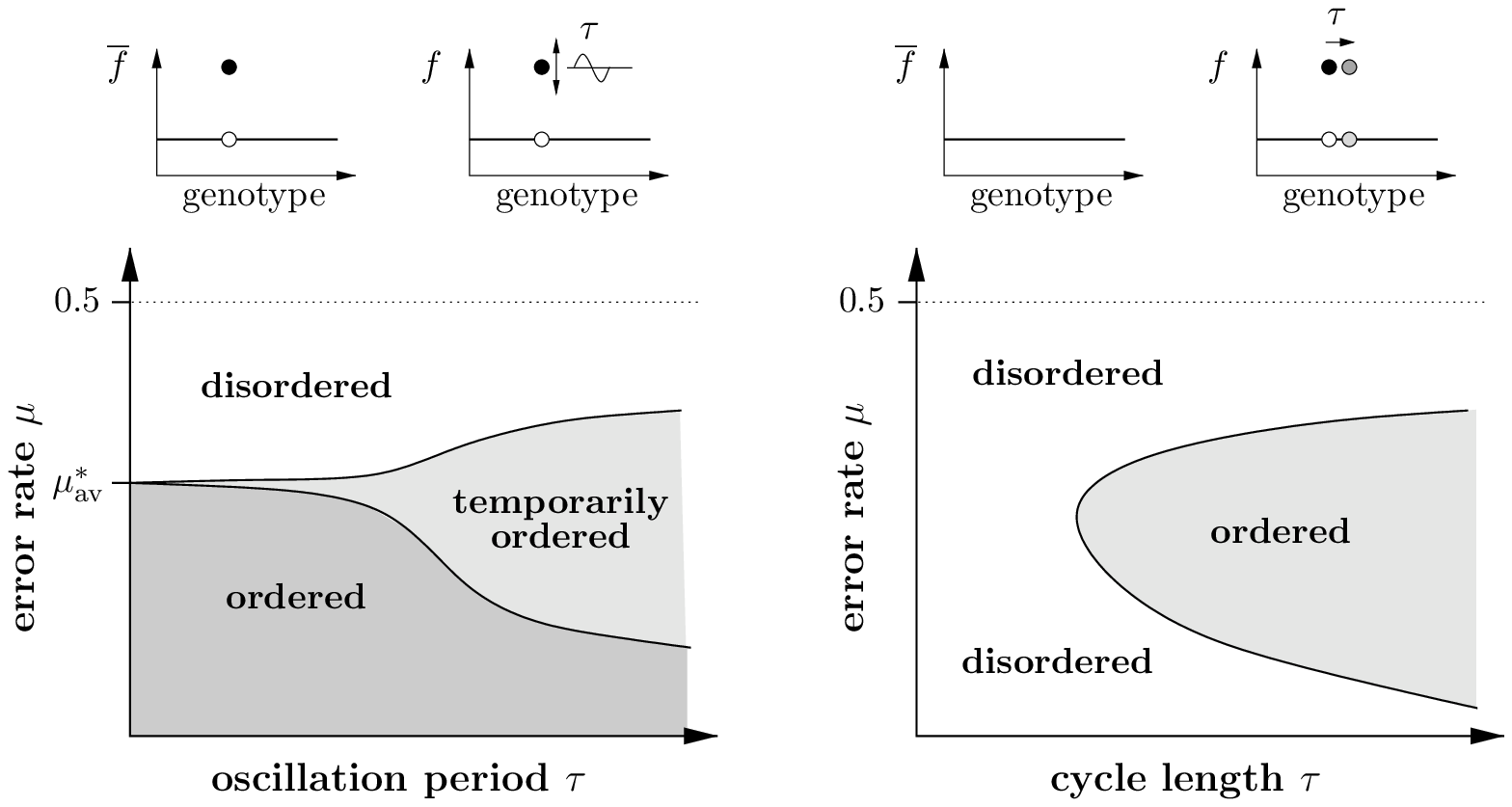}
    \mycaption{%
Phase diagrams for {\it (left)\/}: needle-in-the-haystack
with oscillating height at frequency $\omega=2\pi/\tau$, {\it
(right)\/}: needle-in-the-haystack that jumps after $\tau$ time
steps to a randomly chosen nearest neighbor.\label{fig:ss-phase}}
  \end{center}
\end{figure}

Wilke {\it et al.\ }raise in \cite{wilke99b} the schematic phase diagram of
the continuous Eigen model, which exhibits the same time-average
phases as that for \ssGA{}s. Their result is in perfect agreement with
two recently, explicitly studied time-dependent landscapes. 
First, Wilke {\it et al.\ }studied in \cite{wilke99a} a
needle-in-the-haystack (\nih{}) landscape with oscillating,
$\tau$-periodic fitness of the needle, i.\ e.\
$$f_{0}(t)>f_{1}=\cdots=f_{2^{l}-1}=1\quad\mbox{and}\quad
f_{0}(t)=\sigma\exp\left\{\varepsilon\sin(2\pi\,t/\tau)\right\}.$$
The continuous model was represented for $\delta t\to 0$ as
Eq.\ \ref{eq:disceig} and the periodic quasispecies
Eq.\ \ref{eq:pqs} was calculated. Figure \ref{fig:ss-phase} {\it
  (left)\/} shows the resulting
phase diagram. For small $\tau$, the error threshold is given by the one
of the time-averaged landscape, whereas for large $\tau$, the error threshold 
oscillates between minimum and maximum values corresponding to
$\min_{t}f_{0}(t)$ and $\max_{t}f_{0}(t)$, as expected in the
quasi-static regime. Second, Nilsson and Snoad studied in \cite{nilsson} a moving
\nih{} that jumps randomly to one of its nearest neighbor strings
every $\tau$ time steps. The time-average of this landscape over
many jump cycles is a totally flat or neutral landscape, which
explains the extension of the disordered phase to small $\mu$ and
small $\tau$ as it is shown in Fig.\ \ref{fig:ss-phase} {\it
  (right)\/}. In the quasi-static region, order is expected because
the needle stays long enough at each position for a quasispecies to grow.
Hence, we can understand the existence of the observed and calculated phase diagrams in
Fig.\ \ref{fig:ss-phase} from simple arguments. In fact, they are
special instances of the general schematic phase diagram depicted in
Fig.\ \ref{fig:phase}.

In the following, we will consider regularly moving
\nih{}s and derive the infinite population
behavior of a \genGA{} in such landscapes. This is interesting, since
\genGA{}s should be considered to adapt faster to changes compared to \ssGA{}s, as
the missing time-average region of \genGA{}s for small $\tau$
suggests. To clarify whether a different phase diagram compared to
Fig.\ \ref{fig:ss-phase}~{\it(right)\/} emerges for \genGA{}s with
moving \nih{}, we will calculate the phase diagram including the
optimal mutation rate that maximizes a lower bound for the
concentration of the needle string in the population.

\begin{figure}
\begin{center}
\includegraphics[width=0.55\textwidth]{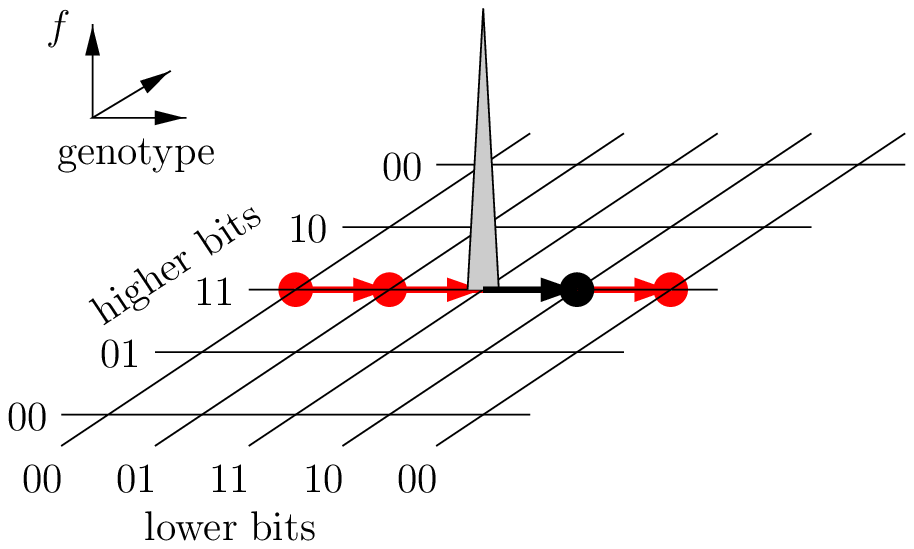}
\qquad\quad\raisebox{0.06\textwidth}{\includegraphics[width=0.25\textwidth]{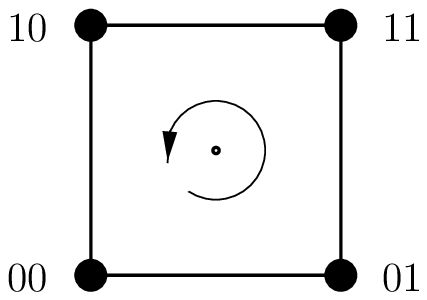}}
\mycaption{%
A regularly moving needle-in-the-haystack for string length $l=4$. 
In {\it (left)\/}, the solid arrow represents
the next jump to happen, whereas the gray and solid arrows all together
represent the jumps that happen one after the other under the rule $P$ of
rotating the two lower bits as shown in {\it (right)\/} with rotation
angle $\pi/2$ at every jump.\label{fig:peak}}
\end{center}
\end{figure}

\section{Generational \GA{} and a moving \nih{}}\label{sec:genga}

In this section, we want to analyze quantitatively the asymptotic behavior of a
\genGA{} with \nih{} that moves
{\it regularly\/}  in the sense of Section \ref{sec:fp} 
to one of its $l$ nearest neighbors every $\tau$ time steps. 
At the end, we will also be able to comment on the case of a \nih{} that jumps
{\it randomly\/} to one of its nearest neighbors. 

\begin{figure}
\begin{center}
\includegraphics[width=0.35\textwidth]{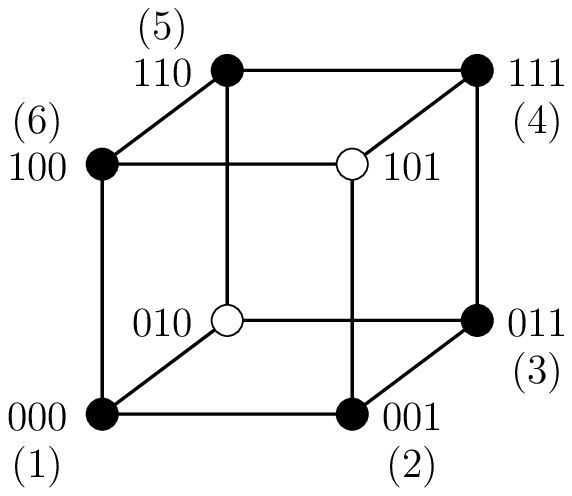}
\qquad\quad\raisebox{0.08\textwidth}{$\begin{array}[b]{c@{\to}c|c@{\to}c}
\multicolumn{4}{c}{P_{\oplus\ll}}\\
\hline\hline
(1)\ 000&001&(4)\ 111&110\\
(2)\ 001&011&(5)\ 110&100\\
(3)\ 011&111&(6)\ 100&000\\
\hline
\phantom{(1)\ }010&101&\phantom{(1)\ }101&010
\end{array}$}
\mycaption{%
The orbit of $0\cdots0$ under $P_{\oplus\ll}$ (black dots) for string
length $l=3$. The numbers $(1),\ldots,(6)$ show the order in which the
strings are visited by the needle, starting from $000$.\label{fig:corb}}
\end{center}
\end{figure}

A simple example of a \nih{} that moves regularly to nearest neighbors is shown in
Fig.\ \ref{fig:peak} {\it(left)\/}.
Each jump corresponds to a $\pi/2$-rotation of the four-dimensional
hypercube $\{0,1\}^{4}$ along the $1100$ axis, i.\ e.\ the lower two
bits are rotated as shown in Fig.\ \ref{fig:peak}~{\it(right)\/}. We
will call the set of strings $\{P^{n}\,i\mid n\in\NN\}$ which is obtained
by applying the same rule $P\in\Perm_{2^{l}}$ over and over to some initial string
$i\in\{0,1\}^{l}$, the {\it orbit of $i$ under $P$\/}. The
period length $4$ of the orbit shown in Fig.\ \ref{fig:peak}~{\it(left)\/} 
originates from the
rotation angle $\pi/2$ and hence is independent of the string length
$l$. The orbits of such rotations will always be restricted to only four
different strings. For reasons that will become clear below, we
are looking for {\it regular\/} movements of the needle that are {\it not\/}
restricted to such a small subspace of the string space. Instead, the needle
is supposed to move `straight away' from previous positions in string
space. Since a complete classification and analysis of all possible {\it regular\/}
movements for given string length $l$ and jump distance $d$ is out of
the scope of this work, we will simply give an example of a rule
$P\in\Perm_{2^{l}}$ that generates such movements: the composition of
a cyclic 1-bit left-shift, which we denote by $P_{\ll}$, and
an exclusive-or with $0\cdots01$, which we denote by $P_{\oplus}$. 
\begin{figure}
\begin{center}
\includegraphics[width=0.35\textwidth]{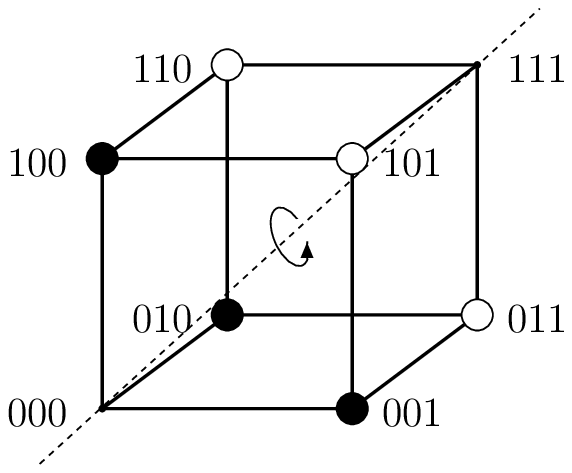}
\qquad\quad\raisebox{0.08\textwidth}{$\begin{array}[b]{c@{\to}c|c@{\to}c}
\multicolumn{4}{c}{P_{\ll}}\\
\hline\hline
000&000&111&111\\\hline
001&010&011&110\\
010&100&110&101\\
100&001&101&011
\end{array}$}
\mycaption{%
The equivalence of a $2\pi/3$-rotation along the $1\cdots1$ axis and a
cyclic 1-bit left-shift, denoted by $P_{\ll}$, for string length
$l=3$.\label{fig:crot}}
\end{center}
\end{figure}
For string length $l\le3$, $P_{\ll}$ corresponds to a $2\pi/l$ rotation
along the $1\cdots1$ axis as can be seen in
Fig.\ \ref{fig:crot}. Moreover, the orbit of $0\cdots0$ under
$P_{\oplus\ll}=P_{\oplus}\circ P_{\ll}$ is shown in
Fig.\ \ref{fig:corb} also for $l=3$. For arbitrary string length $l$,
it is more difficult to visualize the action of $P_{\ll}$ and hence of
$P_{\oplus\ll}$. But, it is
easily verified that starting from all zeros $0\cdots0$, the string
with $n\le l$ ones $0\cdots01\cdots1$ will be reached after exactly
$n$ jumps. Moreover, the orbit of $0\cdots0$ under $P_{\oplus\ll}$ has
the period length $2l$. In the limit of
long strings $l\to\infty$, this periodicity is broken because the needle
never (i.\ e.\ after $\infty$ many jumps) returns to all zeros
$0\cdots0$, but -- as we have shown in
Eq.\ \ref{eq:pqs} using Eq.\ \ref{eq:xxp} -- there still exists an
asymptotic quasispecies.

\begin{figure}
\hskip-0.08\textwidth\includegraphics[width=1.2\textwidth]{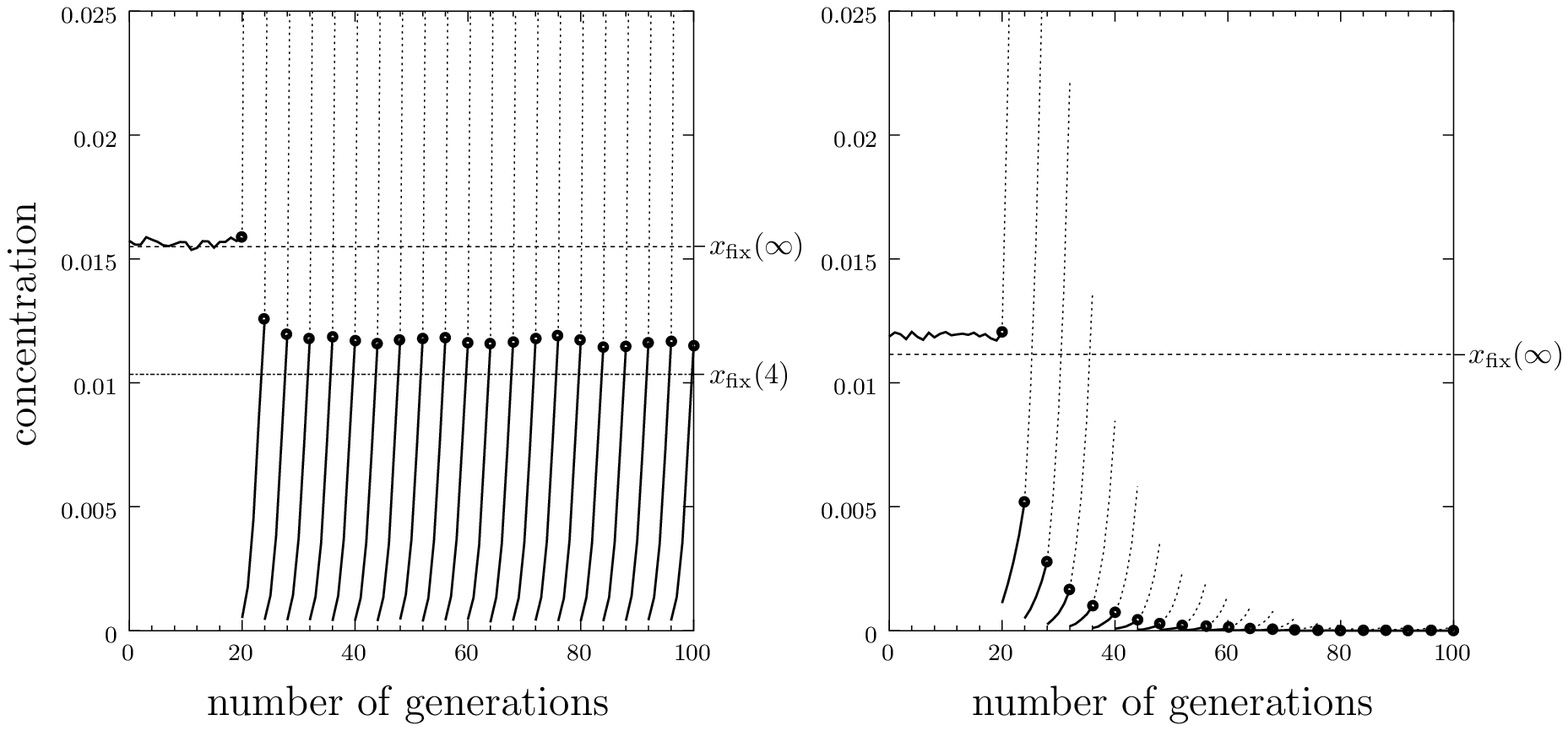}
\begin{center}
\mycaption{%
Run of a \genGA{} with {\it regularly\/} moving
needle-in-the-haystack. The parameter setting was $m=1000000, l=20,
f_{0}=5, \tau=4,$ {\it (left):\/} $\mu=0.022$, {\it (right):\/} $\mu=0.055$. 
In both cases the system evolved for 100 generations (not shown) without any
occurring jumps in order to let a typical quasispecies grow around the
initial needle string. In generation $20$ the first jump happened and
afterwards every $\tau=4$ generations. solid line: $x_{1}(n,t)$,
dotted line: $x_{0}(n,t)$, bullet: jump -- $x_{0}(n+1,0)=x_{1}(n,\tau)$.
\label{fig:regular}}
\end{center}
\end{figure}

How does our simple \GA{} behave with a \nih{} that moves according to
$P_{\oplus\ll}$? In
Fig.\ \ref{fig:regular}, two typical runs of a \genGA{} with a \nih{}
like that are depicted. The setting
$(m,l,f_{0},\tau)$ was kept fixed but two different mutation rates
$\mu$ were chosen. In the case of
Fig.\ \ref{fig:regular}~{\it(right)\/}, the mutation rate is `too
high' to allow the population to track the movement. The concentration of
the future needle string (solid line) cannot grow much within one jump
cycle resulting in a decreasing initial condition (bullet) for the
growth of the needle concentration (dotted line) in the next
cycle. The population looses the peak -- in this case after $\approx 90$
generations. It might happen that the population finds the needle
again by chance (or better saying the moving needle jumps into the
population), but the population will not be able to stably track the
movement. Contrasting to that, the mutation rate was chosen to
maximize the concentration of the future needle string at the end of
each jump cycle (bullets) in Fig.\ \ref{fig:regular}~{\it(left)\/}. 
Since in that case, the best achievable initial condition is given to each jump
cycle, the movement of the needle is tracked with the highest possible
stability for the given setting $(m,l,f_{0},\tau)$. As can be expected
from Fig.\ \ref{fig:regular} and is affirmed by further experiments,
the bullets keep on fluctuating around an average value for
$n\to\infty$ which is for the infinite population given by the
quasispecies Eq.\ \ref{eq:pqs}.
In the following, we are going to model that system with
some idealizations and we will calculate a lower boundary for this
average value.

We adopt the viewpoint of permuting the concentration vector
compatible to the movement of the needle as we have done implicitly in
Fig.\ \ref{fig:regular} and formally in the definition of $\xx'(t)$ 
in Eq.\ \ref{eq:xxp}, but we drop the primes henceforth. 
The concentration of the
needle string within jump cycle $n$ is denoted by $x_{0}(n,\pha)$ and
the concentration of the string the needle will move to with the
$(n+1)$th jump (i.\ e.\ the future needle string in jump cycle $n$) 
is denoted by $x_{1}(n,\pha)$.
The initial cycle prior to which {\it no\/}
jump has occurred is $n=0$. Within a cycle, the time or generation is
counted as phase $\pha\in\{0,\ldots,\tau\}$.
Two succeeding cycles are connected by the (approximated) rule of change
\begin{equation}\label{eq:approx}
x_{0}(n+1,0)=x_{1}(n,\tau)\quad\mbox{and}\quad x_{1}(n+1,0)\approx 0.
\end{equation}
The second relation is an approximation which is made to simplify the
coming calculations, but it holds only if the needle
jumps onto a string which has not been close to one of the previous
needle positions. Otherwise, the future needle string could already be
present with a concentration significantly larger than
$1/2^{l}\approx0$. In Fig.\ \ref{fig:regular}, we have chosen the rule 
$P_{\oplus\ll}$ to get experimental data for a case in which this
assumption is fulfilled. Later on we will see that we can still
make useful comments about cases in which that approximation is
partly broken.

\begin{figure}
\begin{center}
\hbox{\hss\includegraphics[width=.8\textwidth]{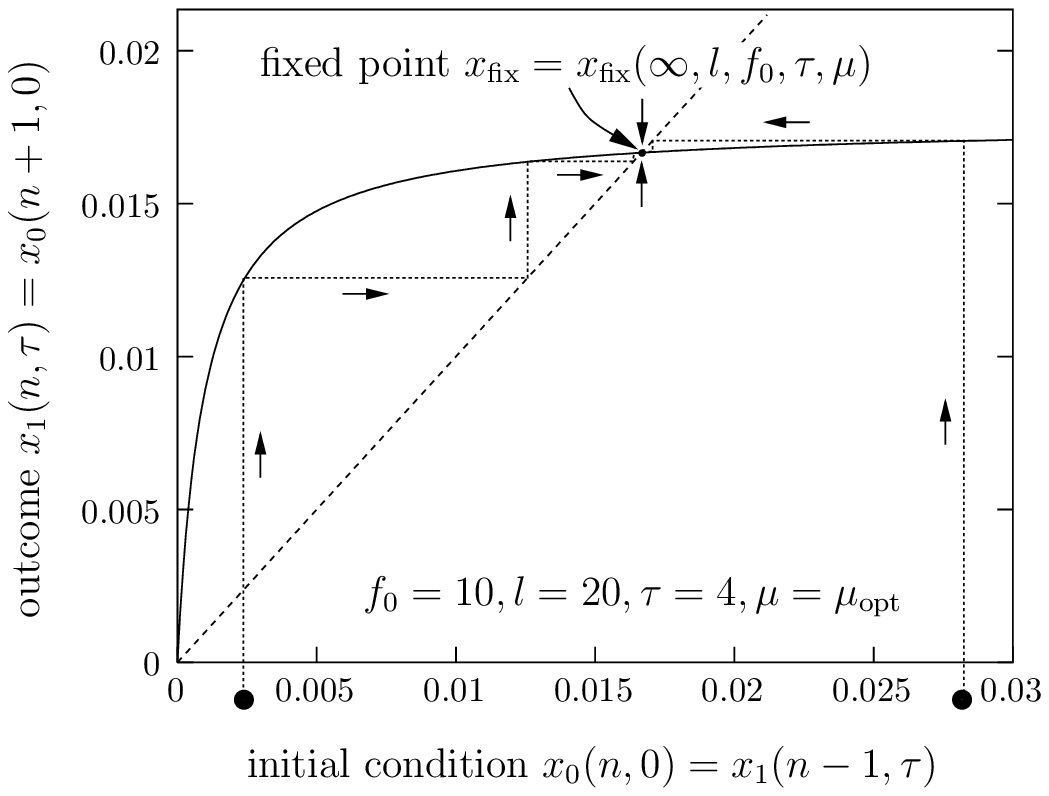}\hss}
\mycaption{%
The fixed point which is reached by an infinite population for
$n\to\infty$.\label{fig:iterfix}}
\end{center}
\end{figure}

If we plot $x_{0}(n+1,0)=x_{1}(n,\tau)$ against $x_{0}(n,0)$, we get
an intuitive picture for the system's evolution towards the
quasispecies. The concentration $x_{0}(n,0)$ converges for
$n\to\infty$ towards a fixed point,
\begin{equation*}
x_{\rm fix}:=\lim_{n\to\infty}x_{0}(n,0),
\end{equation*}
as shown in Fig.\ \ref{fig:iterfix} for a finite value of $x_{\rm fix}$.
Obviously, this fixed point depends on the full setting 
$x_{\rm fix}=x_{\rm fix}(m,l,f_{0},\tau,\mu)$. Since we are especially
interested in the effects of various cycle lengths $\tau$ and mutation
rates $\mu$, we keep $(m,l,f_{0})$ fixed, such that 
$x_{\rm fix}=x_{\rm fix}(\tau,\mu)$.

In the remaining of this section, we will calculate
$x_{0}(n+1,0)=x_{1}(n,\tau)$ 
in dependence on $x_{0}(n,0)$, which is the solid curve in 
Fig.\ \ref{fig:iterfix}, for arbitrary parameter settings. 
From this knowledge, we will construct
the phase diagram. Since we stay within one jump cycle,  
we drop $n$ to take off some notational load.

\subsection{Derivation of the Fixed Point Concentrations}
To calculate $x_{1}(\tau)$, it is sufficient to take only $x_{0}$ and
 $x_{1}$ into account, because the assumed initial condition is
 $x_{1}(0)\approx 0$, such that the main growth of $x_{1}$ is produced
 by the mutational flow from the needle. Moreover, we
 assume $\mu$ to be small
enough such that terms proportional to $\mu^{2}$ can be
neglected. This means we restrict ourselves to the case in which
the system is mainly driven by one-bit mutations. Without
normalization, the evolution equations then read
\begin{equation}\begin{aligned}
y_{0}(t+1)&=\phantom{\mu}(1-\mu)^{l\phantom{-1}}f_{0}\,y_{0}(t)
+\big\{\mu(1-\mu)^{l-1}\,y_{1}(t)\big\},\\
y_{1}(t+1)&=\mu(1-\mu)^{l-1}f_{0}\,y_{0}(t)
+\phantom{\big\{\mu}(1-\mu)^{l\phantom{-1}}\,y_{1}(t),
\label{eq:tevol}\end{aligned}
\end{equation}
where $y_{i}$ denote unnormalized concentrations in contrast to the
normalized concentrations $x_{i}$.

For $f_{0}(1-\mu)\gg \mu$, which is always the case for large enough $f_{0}$, 
we can further neglect the back-flow $\{\cdots\}$ from the future
needle string compared to the self-replication of the current needle
string. The solution
of Eq.\ \ref{eq:tevol} is then given by
\begin{equation*}
\begin{aligned}
y_{0}(t)&=\left[(1-\mu)^{l}f_{0}\right]^{t}y_{0}(0),\\
y_{1}(t)&=\kappa_{t}(\mu)\,y_{0}(0)+(1-\mu)^{lt}y_{1}(0),\\[1.5ex]
&\hskip2cm\mbox{with $\left\{\begin{aligned}
\kappa_{t}(\mu)&=\mu(1-\mu)^{lt-1}\alpha_{t}\\
\alpha_{t}&=\textstyle\sum_{\nu=1}^{t}f_{0}^{\nu}=f_{0}
\frac{f_{0}^{t}-1}{f_{0}-1}.
\end{aligned}\right.$}
\end{aligned}
\end{equation*}
The coefficient $\kappa_{t}(\mu)$ measures the growth of $y_{1}(t)$
starting from the initial condition $y_{1}(0)\approx0,
y_{0}(0)\not=0$. As long as $y_{0}(t)+y_{1}(t)\ll1$, this gives already
a good approximation for the concentrations $x_{0}(t)$ and $x_{1}(t)$. But
in general, this approximation breaks down for large $t$, because of
the exponential growth of $y_{0}(t)$. 
We need to normalize our solution, which can be done by
\begin{equation}
\xx(t)=\yy(t)\big/\langle f\rangle_{0}\cdots\langle f\rangle_{t-1},
\quad\mbox{where $\langle f\rangle_{t}=(f_{0}-1)x_{0}(t)+1$.}
\label{eq:norm}\end{equation}
By expressing the fitness averages in terms of $y_{0}(t)$, we find, after
solving a simple recursion,
\begin{equation*}
\begin{aligned}
\langle f\rangle_{0}\cdots\langle f\rangle_{t-1}&= \textstyle 1+(f_{0}-1)\left[
\sum_{\nu=0}^{t-1}(1-\mu)^{l\nu}f_{0}^{\nu}\right]x_{0}(0)\\
&=1+(f_{0}-1)\beta_{t}(\mu)x_{0}(0),\\[1.5ex]
&\hskip2cm\mbox{where
  $\beta_{t}(\mu)=\frac{\tilde{f}^{t}-1}{\tilde{f}-1}\mbox{ and }
  \tilde{f}=(1-\mu)^{l}f_{0}$}.
\end{aligned}
\end{equation*}
Finally, we arrive at the normalized concentrations
\begin{equation*}
\begin{aligned}
x_{0}(t)&=\left[(1-\mu)^{l}f_{0}\right]^{t}x_{0}(0)\Big/
\left[1+(f_{0}-1)\beta_{t}(\mu)x_{0}(0)\right],\\
x_{1}(t)&=\left[\kappa_{t}(\mu)\,x_{0}(0)+(1-\mu)^{lt}x_{1}(0)\right]\Big/
\left[1+(f_{0}-1)\beta_{t}(\mu)x_{0}(0)\right].
\end{aligned}
\end{equation*}
The asymptotic state can now be calculated by using the initial
condition $x_{1}(0)\approx 0, x_{0}(0)\not=0$ and demanding
$x_{1}(\tau)=x_{0}(0)$. It is easily verified that for the fixed
point follows
\begin{equation}
x_{\rm fix}(\tau, \mu)=\frac{\kappa_{\tau}(\mu)-1}{(f_{0}-1)\beta_{\tau}(\mu)}.
\label{eq:fix}\end{equation}

\begin{figure}
\begin{center}
\hbox{\hss\includegraphics[width=.9\textwidth]{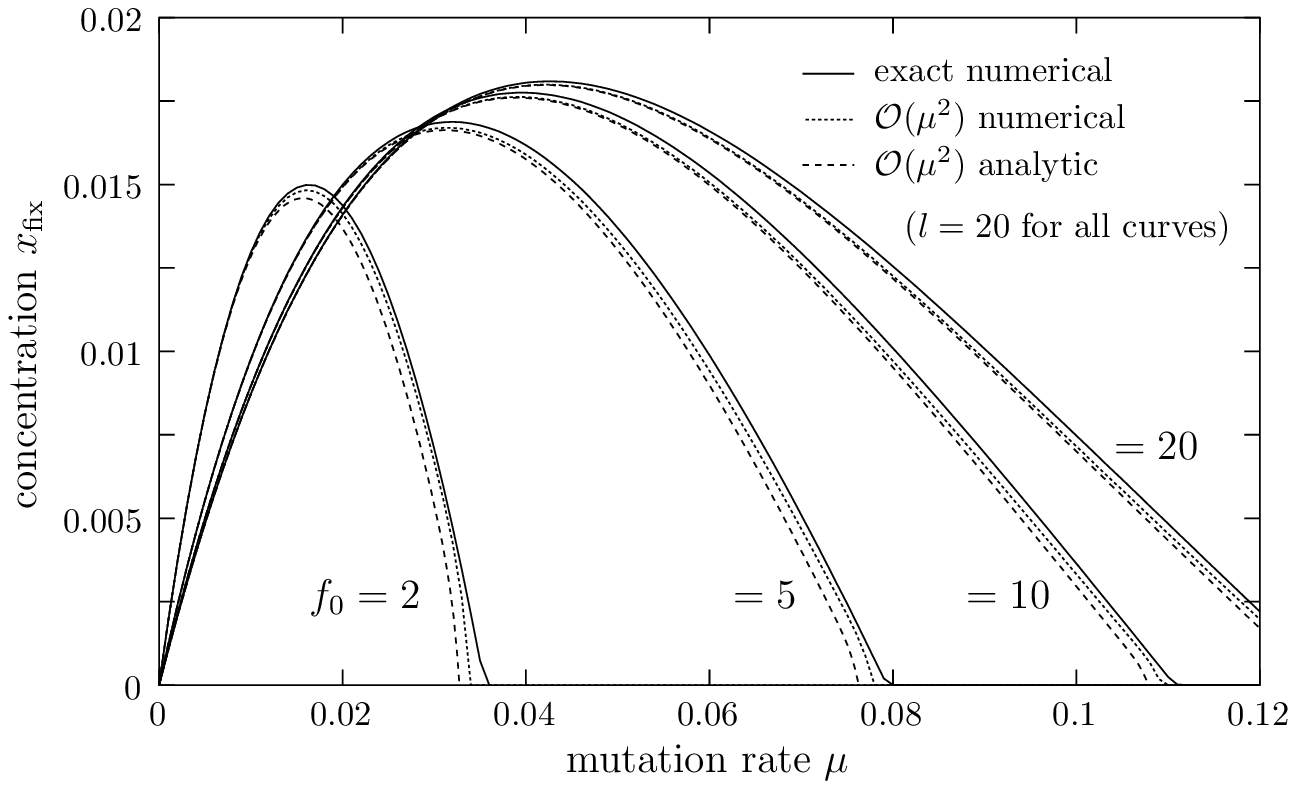}\hss}
\mycaption{%
Comparison of the exact numerical and the $\mathcal{O}(\mu^{2})$
calculation for different values of the needle fitness $f_{0}$.\label{fig:approx}}
\end{center}
\end{figure}

\subsection{Consistency in the Quasi-Static Limit}\label{sec:cons}
How can we test the quality of the approximate result Eq.\ \ref{eq:fix}? For large
cycle lengths $\tau$, we enter the quasi-static regime, where we can
approximate the population at the end of each cycle by the
quasispecies of the corresponding static landscape. 
Figure \ref{fig:approx} shows
a comparison of the exact numerical calculations of the quasispecies
($\tau\to\infty$) and the $\mathcal{O}(\mu^{2})$ calculations
($\tau=100$). In the numerical $\mathcal{O}(\mu^{2})$ calculation, the
back-flow from the first error class to the needle string is
included. Overall, we find the error threshold and the maximum of the
fixed point concentration well represented. This also suggests
that the deviation of the $\mathcal{O}(\mu^{2})$ approximation from
the exact values should be small for smaller $\tau$, because
those deviations add up for $\tau\to\infty$ by the iterative
procedure. 

How do the calculated fixed point concentrations compare to
simulations with (large) finite population? In Fig.\ \ref{fig:regular}, the
values of $x_{\rm fix}(\infty,\mu)$ and $x_{\rm fix}(4,\mu)$ are
shown. For $\tau\to\infty$, the deviation from the average $\langle
x_{1}(n,\pha)\rangle$ (in generations $0-20$) is in fact the same as
what can be read off in Fig.\ \ref{fig:approx}. The deviation of $x_{\rm
  fix}(4,\mu)$ from the average value $\langle x_{0}(n,0)\rangle$ in
generations $24,28,\ldots,100$ is significantly larger. This is caused
by the neglect of all other strings' contributions apart from
the current needle string's contribution to the flow onto the future needle string. 
These neglected contributions increase
the average fixed point concentration measured in the experiment
in comparison to the calculated value $x_{\rm fix}(\tau,\mu)$. 
But even though there are deviations, we conclude that
the approximately calculated value is always a lower bound for the
exact value. In the next section, we will use this observation to
derive an expression for the mutation rate that maximizes the average
fixed point concentration.

\begin{figure}
\begin{center}
\hbox{\hss\includegraphics[width=.9\textwidth]{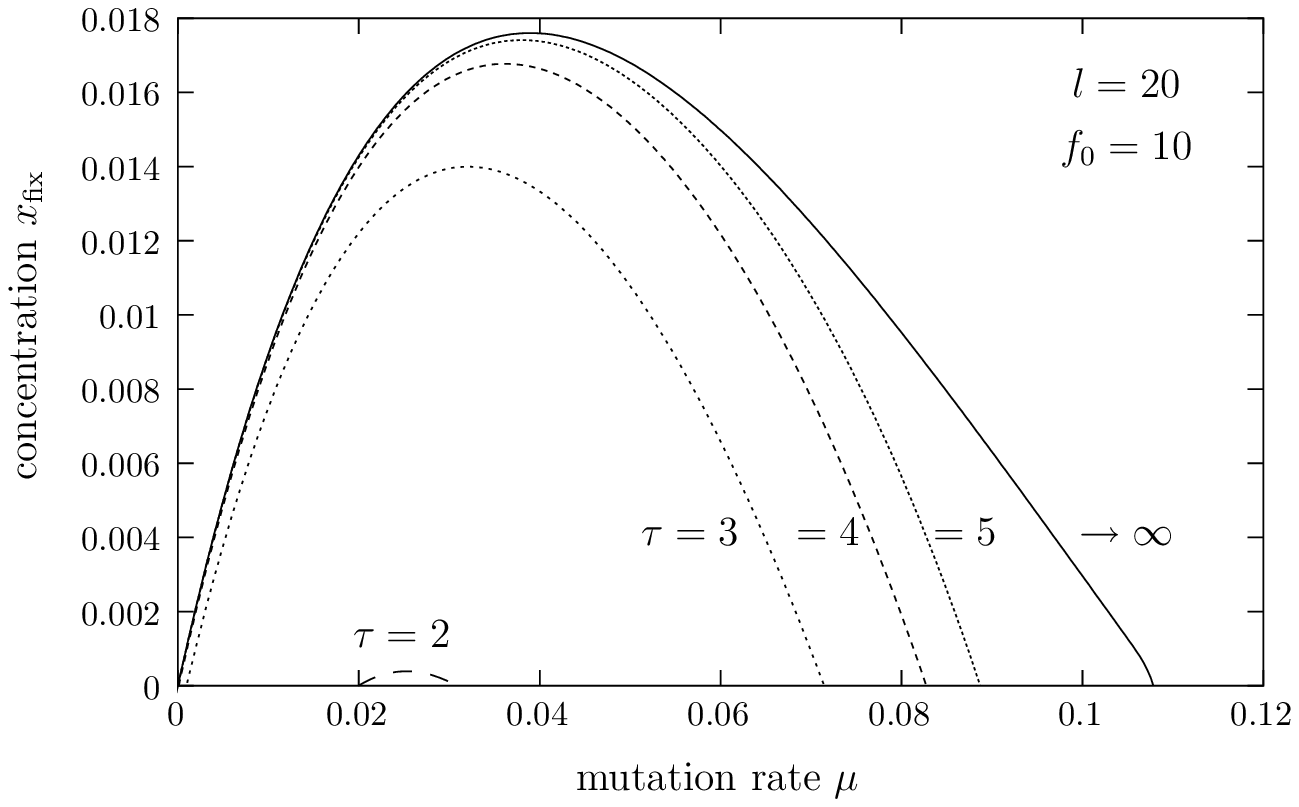}\hss}
\mycaption{%
Fixed point concentration $x_{\rm fix}(\tau,\mu)$ for different values of
$\tau$. For faster changes, the fixed point concentration rapidly
drops down.\label{fig:taudep}}
\end{center}
\end{figure}

\subsection{Phase Diagram}\label{sec:pd}
In Fig.\ \ref{fig:taudep}, the fixed point values $x_{\rm fix}(\tau,\mu)$ are
shown for small cycle lengths $\tau$. For the shown parameter setting,
the region with $x_{\rm fix}(2,\mu)>0$ is extremely small. We
notice that there are two error thresholds, one for `too low' mutation
rates, $\mu_{{\rm th}<}$, and one for `too high' mutation rates, $\mu_{{\rm th}>}$. The
intuition behind that was already given in Section~\ref{sec:spd}.
For too low mutation rates the population becomes
slow and evolves in the averaged, flat landscape, whereas for too high
mutation rates the usual transition to the disordered phase takes
place. In the following we will calculate the phase diagram starting from
Eq.\ \ref{eq:fix}.

\paragraph{Error Thresholds:}
 The error thresholds are given by
\begin{equation}
x_{\rm fix}(\tau,\mu)=0\quad\Longleftrightarrow\quad
\kappa_{\tau}(\mu)=1.
\label{eq:et}\end{equation}
This is the same condition as one would get using only unnormalized concentrations
$y_{i}(t)$. Since $y_{i}(t)\approx0$ near the error thresholds, the
neglect of the normalization is not critical for the calculation of
the error thresholds themselves, whereas it is important for the
optimal mutation rate and of course for the fixed point concentration. 
Since Eq.\ \ref{eq:et} cannot be solved for $\mu$ in closed form, we
write down the corresponding recursion relation that converges, for a
suitable starting value of $\mu$, to the solution of Eq.\ \ref{eq:et} in
the limit $k\to\infty$,
\begin{equation*}\begin{aligned}
\mu_{{\rm th}<}^{(k)}&=1\Big/\alpha_{\tau}\left(1-\mu_{{\rm th}<}^{(k-1)}\right), &
\mu_{{\rm th}<}^{(0)}&=0,\\
\mu_{{\rm th}>}^{(k)}&=1-\left(1\Big/\alpha_{\tau}\mu_{{\rm th}>}^{(k-1)}\right)^{1/(l\tau-1)},
&\quad \mu_{{\rm th}>}^{(0)}&=1-f_{0}^{-1/l}=:\mu_{{\rm th}}^{\infty}.
\end{aligned}\end{equation*}
For $\mu_{{\rm th}<}$, a good starting value is $0$, since $\mu_{{\rm th}<}\approx0$
anyway. For $\mu_{{\rm th}>}$, the approximate value for the
error threshold of the static (i.\ e.\ $\tau\to\infty$)
landscape $\mu_{{\rm th}}^{\infty}$ can be chosen, which is obtained by
calculating the fixed point [using Eq.\ \ref{eq:tevol} and \ref{eq:norm}],
$$x_{0}(t+1)=x_{0}(t)\quad\Longleftrightarrow\quad 
x_{\rm fix}^{\infty}=\frac{(1-\mu)^{l}f_{0}-1}{f_{0}-1},$$
setting it to zero and solving for $\mu$.

\paragraph{Optimal Mutation Rate:} In order to track changes with the
best achievable stability for a given setting $(m,l,f_{0},\tau)$,  
the lowest possible concentration (infimum of)
$x_{0}(n,\pha)$ needs to be maximized, because a low concentration
might result in the loss of the needle string in a finite
population. Since for infinite populations $x_{0}(n,\pha)$ is monotonously
increasing with $\pha$ it is sufficient to maximize $x_{0}(n,0)$. 
Moreover, we derived above that $x_{0}(n,0)$ approaches the fixed
point value $x_{\rm fix}(\tau,\mu)$ for $n\to\infty$. For finite
populations, we expect similar behavior but the strict monotony of
$x_{0}(x,\pha)$ in $\pha$ will be destroyed by fluctuations and also the fixed
point value itself will fluctuate around some average value $\langle
x_{\rm fix}\rangle$ as can be seen in Fig.\ \ref{fig:regular}. 
However, the {\it safest\/} way to avoid any loss of the needle string
is still to maximize the average fixed point value $\langle x_{\rm fix}\rangle$. 
In this sense, we define the {\it optimal mutation rate\/} $\mu_{\rm opt}$ 
as the one that maximizes $\langle x_{\rm fix}\rangle$.
In the previous Section~\ref{sec:cons}, we noted that our approximated
infinite population value $x_{\rm fix}(\tau,\mu)$ represents a lower bound for $\langle
x_{\rm fix}\rangle$, where the maxima of the two curves are expected
to coincide for fixed $\tau$. Thus, $\mu_{\rm opt}$ can be obtained by
maximization of $x_{\rm fix}(\tau,\mu)$.  

We can derive an expression for the optimal mutation rate $\mu_{\rm opt}$ from
$$\frac{\partial x_{\rm fix}}{\partial \mu}(\tau,\mu_{\rm opt})=0$$
If we neglect the $\mu$ dependence of $\beta_{\tau}(\mu)$ in
Eq.\ \ref{eq:fix}, which corresponds to the approach in \cite{nilsson}, 
we simply find $\mu^{\rm NS}_{\rm opt}(\tau,l)=1/l\tau$. Because of
$\smash{\mu^{\rm NS}_{\rm opt}\xrightarrow{\tau\to\infty} 0}$, this result is
inconsistent with the quasi-static limit, because $\mu_{\rm opt}$ should
approach the value for which the concentration of 1-mutants in the
quasispecies of the corresponding static \nih{} landscape is maximized. 
We conclude that the $\mu$ dependence of
$\beta_{\tau}(\mu)$ cannot be neglected for the correct
optimal mutation rate, which we are now going to calculate. 

\begin{figure}
\begin{center}
\hbox{\hss\includegraphics[width=.9\textwidth]{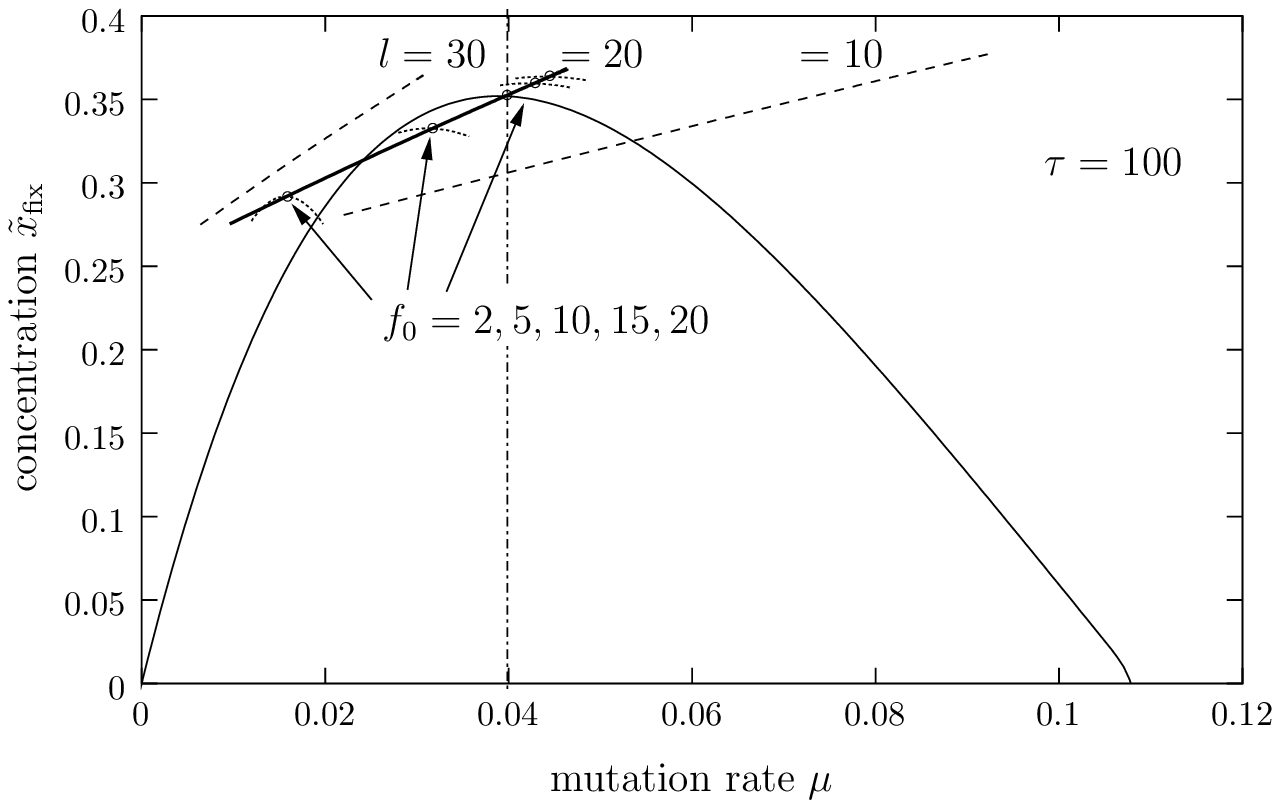}\hss}
\mycaption{%
The optimal mutation rate $\mu_{\rm opt}^{\infty}(f_{0},l)$ from
Eq.\ \ref{eq:optmut} in dependence on needle height $f_{0}$ and string
length $l$.\label{fig:optmut}}
\end{center}
\end{figure}

For $\alpha_{\tau}\gg1$, which is the case for $\tau\gg 1$ and
$f_{0}>1$, or $\tau\approx 1$ and $f_{0}\gg 1$, we can neglect the
$-1$ in the numerator of $x_{\rm fix}(\tau,\mu)$ and take only
$\alpha_{\tau}$ into account for the calculation of $\partial x_{\rm fix}
/\partial \mu$. After some
algebra, we find
\begin{equation*}
\mu_{\rm opt}=\frac{(\tilde{f}^{\tau}-1)(\tilde{f}-1)}{l(\tilde{f}^{\tau+1}
-(\tau+1)\tilde{f}+\tau)},\quad\mbox{where $\tilde{f}=f_{0}(1-\mu_{\rm opt})^{l}$}.
\end{equation*}
Since $\tilde{f}=\tilde{f}(\mu_{\rm opt})$, this equation cannot be
solved in a closed form for $\mu_{\rm opt}$. However, for $\tau\to\infty$
the equation simplifies to 
\begin{equation*}
\mu^{\infty}_{\rm opt}=\begin{cases}
\displaystyle(\tilde{f}-1)/l\tilde{f}&: \tilde{f}>1\\
\phantom{f}0&:\tilde{f}\le 1.
\end{cases}
\end{equation*}
In the case $\tilde{f}>1$, we find
\begin{equation*}
(1-l\mu^{\infty}_{\rm opt})(1-\mu^{\infty}_{\rm opt})^{l}=1\big/f_{0}.
\end{equation*}
By approximating $(1-\mu)^{l}\approx (1-l\mu)^{2}$, we get a cubic
equation. The real root of that equation is approximately \cite{ropt}
given by (see also Fig.\ \ref{fig:optmut})
\begin{multline}
\mu_{\rm opt}^{\infty}(f_{0},l)\approx \mu_{+}
\left[1+\frac{(l-1)\mu_{+}(1-l\mu_{+})}{3l(l-1)\mu_{+}^{2}-2\mu_{+}(3l-1)+4}
\right]\\[1.5ex]
\mbox{with $\mu_{+}=\displaystyle\frac{1}{l}\left[1+f_{0}^{-1/2}\right]$.}
\label{eq:optmut}\end{multline}

\begin{figure}
\begin{center}
\hbox{\hss\ \hskip0.05\textwidth\includegraphics[width=\textwidth]{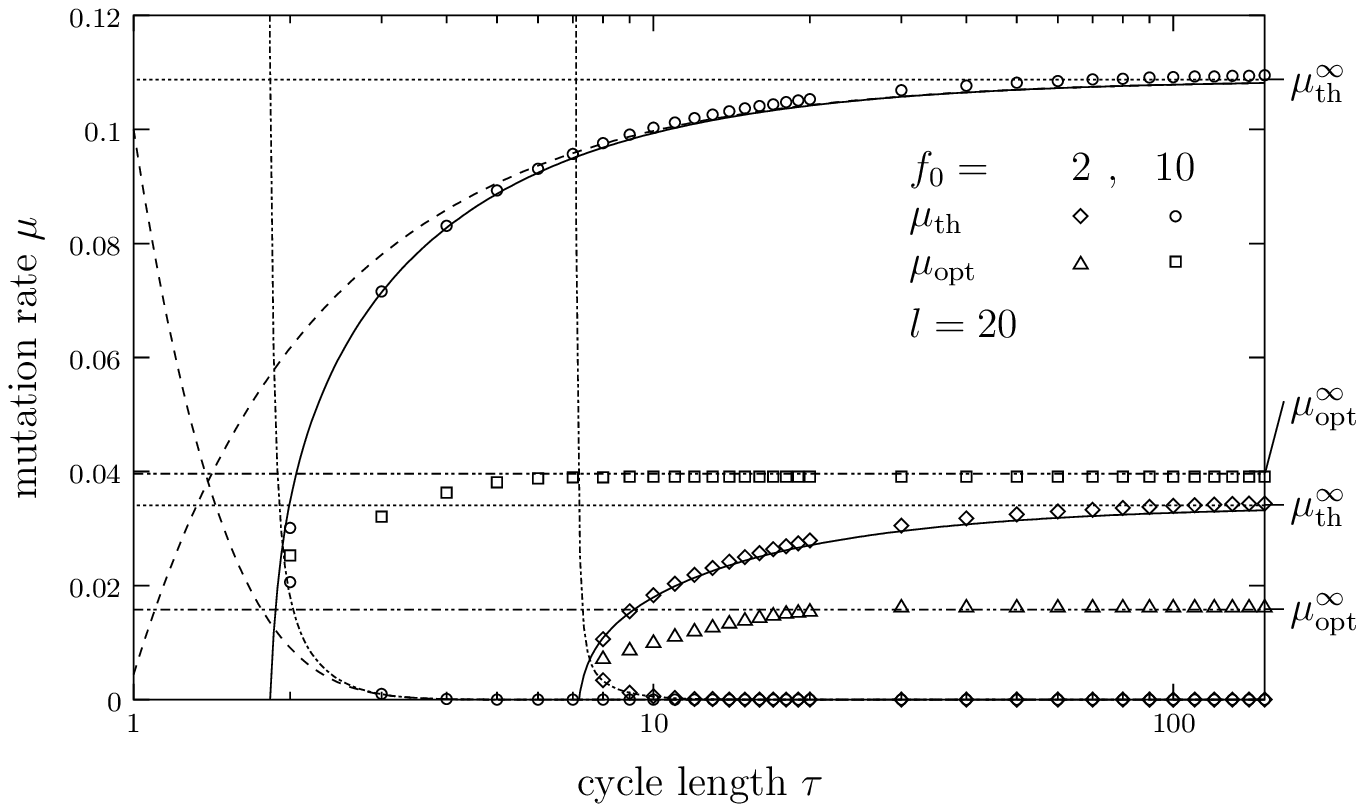}\hss}
\mycaption{%
The calculated phase diagram for a \genGA{} with stochastically moving
needle-in-the-haystack; two settings are shown: $f_{0}=2,10$, for both
$l=20$.\label{fig:pd-genga}}
\end{center}
\end{figure}

\newpage
\paragraph{Resulting Phase Diagram:}
From the above, we are able to plot the phase diagram for our model
as shown in Fig.\ \ref{fig:pd-genga}. Two settings are
plotted. For $f_{0}=2\ (\mbox{resp.\ }10)$ the diamonds (resp.\ circles) are the
numerically obtained error thresholds. The solid and dash-dotted lines
are $\smash{\mu_{{\rm th}<}^{\smash{(5)}}}$ and $\smash{\mu_{{\rm th}>}^{\smash{(5)}}}$. 
To show the convergence
property of $\smash{\mu_{{\rm th}<,>}^{\smash{(k)}}}$, 
$\smash{\mu_{{\rm th}<,>}^{\smash{(0)}}}$ are plotted for
$f_{0}=10$ as dashed lines. Obviously, the needed corrections to the chosen
starting values increase for smaller $\tau$, such that more iterations
are needed to describe the error thresholds correctly for
small $\tau$. The expressions $\smash{\mu^{\smash{(5)}}_{{\rm th}<,>}}$ 
are already a good approximation
for the given settings. Representing the quasi-static limit, $\mu_{\rm
  th}^{\infty}$ is plotted as dotted line and gets consistently
approached by $\mu_{{\rm th}>}(\tau)$ for $\tau\to\infty$. Furthermore,
$\mu_{\rm opt}^{\infty}$ is plotted as dash-dot-dotted line. The
numerically measured values for $\mu_{\rm opt}(\tau)$ are shown for
$f_{0}=2\ (\mbox{resp.\ }10)$ as triangle (resp.\ squares). They approach 
$\mu_{\rm opt}^{\infty}$ very quickly already for $\tau\approx 20\
(\mbox{resp.\ }10)$.

We conclude that the above quantitative description is in good
agreement with the numerical observations and approaches the
quasi-static region in a consistent way. Moreover, the phase diagram
fits well into the general one raised in Section \ref{sec:spd}. 
Even in the considered case of a \genGA{}, we find -- depending on
the parameter setting -- a time-averaged phase for very small
$\tau$. The time-averaged phase broadens for small $f_{0}$.

\subsection{Stochastically moving \nih{}}
\begin{figure}
\begin{center}
\includegraphics[width=0.55\textwidth]{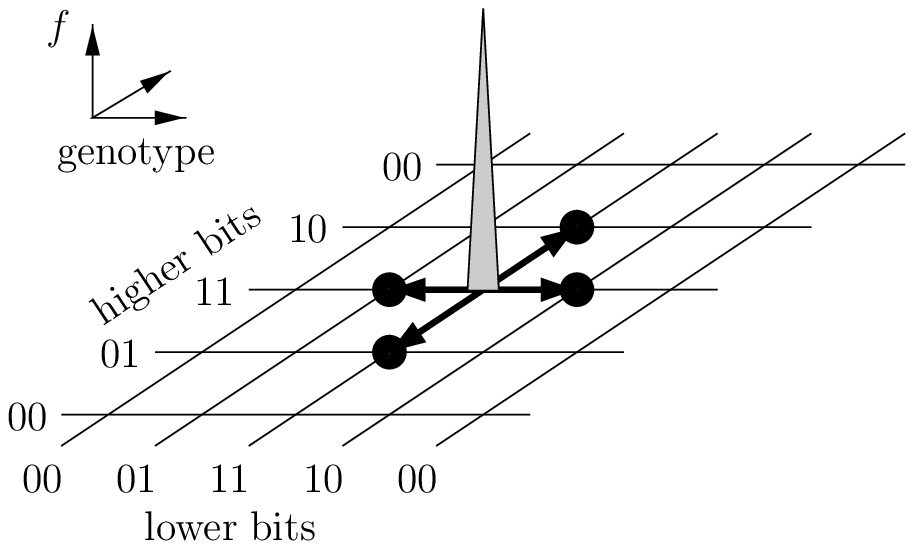}
\mycaption{%
A {\it stochastically\/} moving needle-in-the-haystack for string length $l=4$. 
The needle is allowed to jump to one of its nearest neighbors which is
chosen at random.\label{fig:stochpeak}}
\end{center}
\end{figure}

Up to now, we analyzed a regularly moving \nih{}, for example with the
rule $P_{\oplus\ll}$. What happens if the \nih{} is allowed to move
to a {\it randomly\/} picked nearest neighbor, as it is shown in
Fig.\ \ref{fig:stochpeak} for $l=4$?
\begin{figure}
\hskip-0.08\textwidth\includegraphics[width=1.2\textwidth]{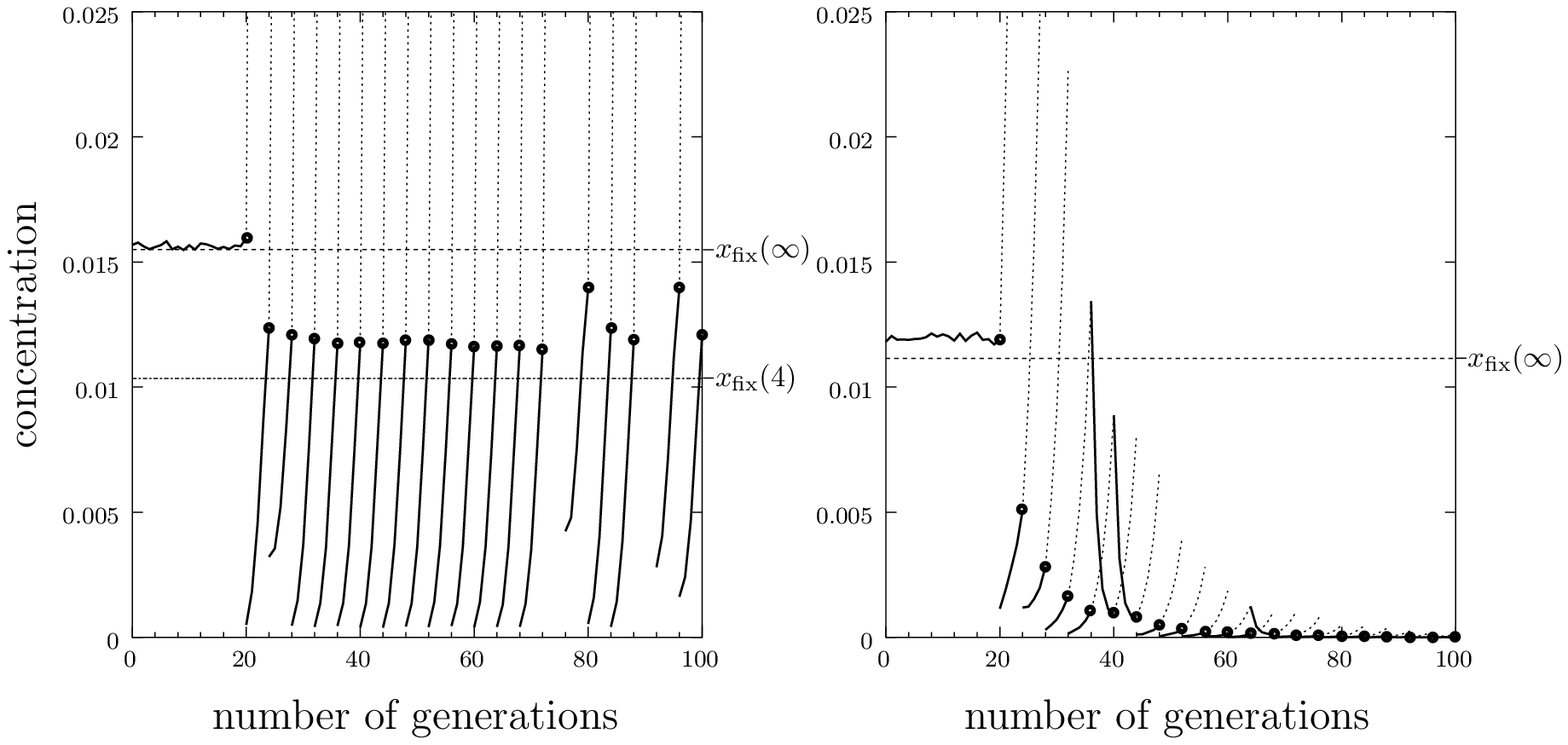}
\begin{center}
\mycaption{%
Run of a \genGA{} with {\it stochastically\/} moving
needle-in-the-haystack. The parameter setting in {\it (left)\/} and {\it
  (right)\/} were the same as in Fig.\ \ref{fig:regular} {\it (left)\/}
and {\it (right)\/}.\label{fig:stoch}}
\end{center}
\end{figure}
Two typical runs of a \genGA{} with this fitness landscape are
depicted in Fig.\ \ref{fig:stoch}. The setting $(m,l,f_{0},\tau)$ was
chosen the same as in Fig.\ \ref{fig:regular} which allows for a
direct comparison of the \GA{}'s behavior for regularly and
stochastically moving \nih{}s. The overall behavior is similar. For
large mutation rates, the population looses the needle string, whereas
the moving needle is tracked stably for mutation rates close to the
above defined optimal mutation rate. In addition, strong fluctuations
in the values of $x_{1}(n,0)$ (lower ends of solid lines) as well as
$x_{0}(n+1,0)=x_{1}(n,\tau)$ (bullets) occur in the {\it stochastic\/}
case. These result from {\it back-jumps\/}. If, at the end of the
current cycle, the needle jumps back to the string it has been to in the
previous cycle, then $x_{1}(n,0)=x_{0}(n-1,\tau)$ is significantly
larger than zero. This can be seen in Fig.\ \ref{fig:stoch} {\it(right)\/}
at generations $36,40$ and $64$ and also in Fig.\ \ref{fig:stoch}
{\it(left)\/} at generations $72$ and $88$ (the gaps in
Fig.\ \ref{fig:stoch} {\it(left)\/} correspond to $x_{1}, x_{0}$ being much
larger than $0.025$). If no back-jumps occur, as in generations
$24-72$ in Fig.\ \ref{fig:stoch} {\it(left)},
the system with stochastic \nih{} behaves nearly indistinguishable
from the one with regularly moving \nih{}. Since back-jumps always increase
the concentrations of the needle string in the very next occurring
jumps, the above calculated fixed point $x_{\rm fix}(\tau,\mu)$ is
still a lower bound. Thus, our previous notion of optimal mutation
rate remains applicable to the stochastically moving \nih{} although
the assumption $x_{1}(n,0)\approx0$ from Eq.\ \ref{eq:approx} is
not always fulfilled. 

Nilsson and Snoad \cite{nilsson} did their analysis of the continuous
Eigen model Eq.\ \ref{eq:conteig} with stochastic \nih{} in a similar way as we did
above. In analogy to their calculation for the continuous Eigen model,
we find for a \genGA{} the optimal mutation rate 
$\mu^{\rm NS}_{\rm opt}(\tau,l)=1/l\tau$ which is inconsistent with
the quasi-static limit (see Section \ref{sec:pd}). The reason is
the missing normalization in the work of Nilsson and
Snoad. Furthermore, they could not derive
an expression for the fixed point concentration $x_{\rm fix}(\tau,\mu)$
because of that same reason.

\subsection{Jumps of larger Distance}
To conclude this section about the behavior of \genGA{}s with
different kinds of \nih{}s
that move to {\it nearest} neighbors, let us shortly discuss jumps of
Hamming distance $d$ {\it larger\/} than one.
Obviously, the analytical calculations get more complicated,
because the $\mathcal{O}(\mu^{2})$-approximation is not sufficient
anymore as it connects only
nearest neighbors. To describe jumps of a larger distance, the
concentrations of some intermediate sequences need to be taken into
account, so that we have to solve a time evolution much more
complicated than Eq.\ \ref{eq:tevol}. Hence, we cannot make
simple statements for finite $\tau$. On the other hand, the system
approaches the quasi-static region for large $\tau$ and it is
characterized by $\mu_{{\rm th}<,>}^{\infty}$ and $\mu_{\rm
  opt}^{\infty}$ as we have seen in Fig.\ \ref{fig:pd-genga}. 
\begin{figure}
\begin{center}
\hbox{\hss\ \hskip0.05\textwidth\includegraphics[width=.9\textwidth]{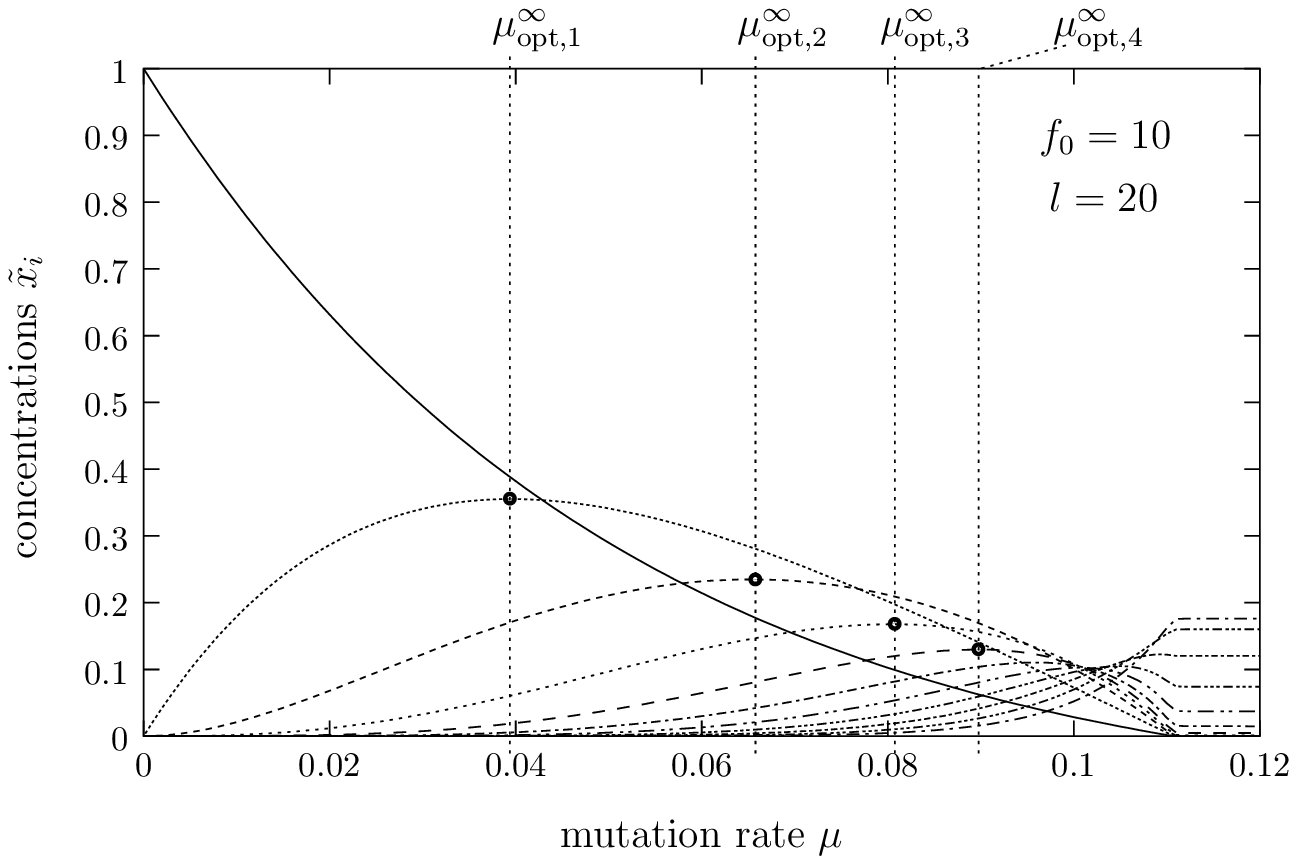}\hss}
\mycaption{%
The quasispecies for the static \nih{} in dependence on the mutation
rate $\mu$. The concentrations $\tilde{x}_{i}$ of the $i$th error class
for $i\in\{0,\ldots,\lfloor l/2\rfloor\}$ are depicted. 
The optimal mutation rates for jumps of Hamming distance
$d=1,2,3,4$ are shown as dotted lines.\label{fig:all-x-r}}
\end{center}
\end{figure}
The exact
quasispecies for $\tau\to\infty$ is shown in Fig.\ \ref{fig:all-x-r}. The plotted
values are error class concentrations, in order to make the higher
error classes visible at all. Each $k$-mutant has a concentration of
$\tilde{x}_{k}/\smash{\left({l \atop k}\right)}$ in the quasispecies state,  
because for a \nih{} the mutant's fitness depends
only on its Hamming distance to the needle and therefore all 
$\smash{\left({l \atop k}\right)}$ $k$-mutants have the same
concentration in the quasispecies. For finite populations, this is
only true on average, because the asymptotic state is distorted by
fluctuations. But in the following, we assume that the quasispecies is still
representative for the average distribution of the population in the
asymptotic state. Then, the optimal mutation rate in the 
sense of Section~\ref{sec:pd} for jumps of distance $d$ is by
definition the position of the maximum of $\tilde{x}_{d}$. 
For $d\ge l/2$, optimal mutation rate and error threshold become identical.
Although $\tilde{x}_{d}$ is maximized for mutation rates close
to the error threshold it amounts, as do all other concentrations
to only $\approx 1/2^{l}$, which leads to an approximately random drift for finite
populations. On the other hand, the chance of tracking the needle
decreases even further for small mutation rates because then the concentration
$\tilde{x}_{d}$ becomes even smaller. In this sense, the quasispecies
distribution, which is centered on the needle
string, is useless for tracking the next jump if $d\ge l/2$.
This also suggests -- in agreement with the
experimental findings of Rowe \cite{roweqs} (in this book) -- that finite populations
are for low mutation rates 
unable to track large jumps -- in particular in the extreme case $d=l$.
Only for jumps of $d<l/2$ the corresponding error class
concentration $\tilde{x}_{d}$ shows a concentration
maximum significantly above $1/2^{l}$. From the heights of the concentration maxima,
we see that the difficulty of tracking the changes increases with
the Hamming distance $d$ of the jumps. Vice versa, the advantage a
population gets after a jump from its structure prior to the jump decreases
with increasing jump distance $d$. 
In addition, a mutation rate which is simultaneously optimal for more
than one distance cannot be found.

\section{Conclusions and Future Work}
On the basis of general arguments, the phase diagrams of
population-based mutation and probabilistic selection systems like the
above \genGA{}, \ssGA{} and Eigen model in time-dependent fitness
landscape can be easily understood. 
The notion of regular changes allows for an exact calculation of the
asymptotic state in the sense of a generalized, time-dependent
quasispecies. For a \genGA{} with \nih{} that moves regularly to
nearest neighbors, the
quasispecies can be straightforwardly calculated under simplifying
assumptions. The result is a lower bound for the exact quasispecies.
With that lower bound, we have constructed the phase diagram in the infinite
population limit. This phase diagram is in agreement with the one
raised from general arguments. 

In order to improve our analysis, we need to weaken our assumptions.
In particular, we have to overcome the restriction of taking into account
only the flow from the current towards the future needle
string. The presence of other contributions to the flow has to be modeled in some way.
Another future step could be an investigation of the fluctuations that are
introduced by the finiteness of realistic populations (discreteness of
$\Lambda_{m}$) around the quasispecies. This would lead to a lower
boundary for the population size above which
the needle string is not lost due to those fluctuations.

An extension of our analysis to non-regularities like the
occurrence of more than a single jump
rule, can be achieved by averaging the time evolution Eq.\ \ref{eq:gtev} 
for $n\to\infty$ according to each rule's probability of being applied. 
A similar averaging procedure will be necessary if fluctuations of the
cycle length $\tau$ are present.
Finally, an extension of the description to broader, more
realistic peaks, as well as \GA{} models including crossover and
other selection schemes, are important topics for future work.


\begin{thebibliography}{xxx}
\bibitem{ropt} A more detailed explanation and analysis of the used approximation
will be presented elsewhere.
\end{thebibliography}
\end{document}